
\magnification 1200
\overfullrule=0pt
\hbadness=10000
\vbadness=10000
\hfuzz=50pt
\vfuzz=50pt
\hsize=16.5truecm \vsize=22truecm \hoffset=.1truecm
\parskip=7pt
\baselineskip=12pt


\let\1\sp


\font\eightrm=cmr8

\def\draftonly#1{\ifx\draft1{\rm \lbrack{\ttraggedright {#1}}
                    \rbrack\quad}                \else\fi}

\def\comment#1{}

\def\today{\ifcase\month\or
 January\or February\or March\or April\or May\or June\or
 July\or August\or September\or October\or November\or December\fi
 \space\number\day, \number\year}

\newcount\refno \refno=1
\def\eat#1{}
\def\labref #1 #2#3\par{\def#2{#1}}
\def\lstref #1 #2#3\par{\edef#2{\number\refno}
                              \advance\refno by1}
\def\txtref #1 #2#3\par{  \def#2{\number\refno
      \global\edef#2{\number\refno}\global\advance\refno by1}}
\def\doref  #1 #2#3\par{{\refno=0  \eat{#2} \if0#2\else
     \vbox {\item {{\bf\lbrack#2\rbrack}} {#3\par}\par}
     \vskip\parskip \fi}}
\let\REF\labref    
\def\Jref#1 "#2"#3 @#4(#5)#6\par{#1:\
    #2,\ {\it#3}\ {\bf#4},\ #6\ (#5)\par}
\def\Bref#1 "#2"#3\par{#1:\ {\it #2},\ #3\par}
\def\Gref#1 "#2"#3\par{#1:\ ``#2'',\ #3\par}

\def\idty{{\leavevmode{\rm 1\ifmmode\mkern -5.4mu\else
                                            \kern -.3em\fi I}}}
\def\Ibb #1{ {\rm I\ifmmode\mkern -3.6mu\else\kern -.2em\fi#1}}
\def\Ird{{\hbox{\kern2pt\vbox{\hrule height0pt depth.4pt width5.7pt
    \hbox{\kern-1pt\sevensy\char"36\kern2pt\char"36} \vskip-.2pt
    \hrule height.4pt depth0pt width6pt}}}}
\def\Irs{{\hbox{\kern2pt\vbox{\hrule height0pt depth.34pt width5pt
       \hbox{\kern-1pt\fivesy\char"36\kern1.6pt\char"36} \vskip -.1pt
       \hrule height .34 pt depth 0pt width 5.1 pt}}}}
\def\Ir{{\mathchoice{\Ird}{\Ird}{\Irs}{\Irs} }}
\def\ibb #1{\leavevmode\hbox{\kern.3em\vrule
     height 1.5ex depth -.1ex width .2pt\kern-.3em\rm#1}}
\def\Nl{{\Ibb N}} \def\Cx {{\ibb C}} \def\Rl {{\Ibb R}}

\def\lessblank{\parskip=5pt \abovedisplayskip=2pt
          \belowdisplayskip=2pt }
\outer\def\iproclaim #1. {\vskip0pt plus10pt \par\noindent
     {\bf #1.\ }\begingroup \interlinepenalty=250\lessblank\sl}
\def\eproclaim{\par\endgroup\vskip0pt plus10pt\noindent}
\def\proof#1{\par\noindent {\it Proof #1}\          
         \begingroup\lessblank\parindent=0pt}
\def\QED {\hfill\endgroup\break
     \line{\hfill{\vrule height 1.8ex width 1.8ex }\quad}
      \vskip 0pt plus100pt}
\def\QEDD {\hfill\break
     \line{\hfill{\vrule height 1.8ex width 1.8ex }\quad}
      \vskip 0pt plus100pt}

\def\nl{\hfill\break}



\def\bra #1 {\langle #1\rangle}

\def\dim {\mathop{\rm dim}\nolimits}

\def\id{\mathop{\rm id}\nolimits}
\def\ker{{\rm ker}\,}
\def\ket #1 {\mid#1\rangle}

\def\mod{{\rm mod}\,}

\def\Order{{\bf O}}
\def\rstr{\hbox{$\vert\mkern-4.8mu\hbox{\rm\`{}}\mkern-3mu$}}

\def\spec{{\rm spec}\,}

\def\tover#1#2{{\textstyle{#1\over #2}}}
\def\tr{\mathop{\rm tr}\nolimits}
\def\Tr{\mathop{\rm Tr}\nolimits}

\def\phi{\varphi}            
\def\epsilon{\varepsilon}    
\def\el{l}

\def\A{{\cal A}} \def\B{{\cal B}}  
\def\G{{\cal G}} \def\HS{{\cal H}} \def\K{{\cal K}} 
\def\M{{\cal M}}   
\def\O{{\cal O}}

\def\E{{\Ibb E}}   \def\F{{\Ibb F}}
 \def\P{{\Ibb P}}

\def\Face{{\cal F}}
\def\el{l}


\def\ie{i.e.\ }               
\def\eg{e.g.\ }               

\def\om{\omega}


\def\es{\vec{{\bf S}}}


\def\boxx#1{\dimen99=#1 \dimen98=\dimen99 \advance \dimen98 by -.8pt
\hbox{\vrule\vtop to \dimen99{\hrule width \dimen98 \vfil
\hrule width \dimen98}\vrule}}

\dimen0=4mm
\dimen9=.4mm
\dimen10=\dimen0 \advance \dimen10 by \dimen9
\def\Ybox{\hbox to \dimen10{\hfil\boxx{\dimen0}\hfil}}

\def\Young #1 #2 #3 #4{
\baselineskip=0pt
\dimen1=\dimen10 \dimen2=\dimen10 \dimen3=\dimen10
\dimen4=\dimen10
\multiply \dimen1 by #1 \multiply \dimen2 by #2 \multiply \dimen3 by #3
\multiply \dimen4 by #4
\advance \dimen1 by .1mm \advance \dimen2 by .1mm \advance \dimen3 by .1mm
\advance \dimen4 by .1mm
\dimen8=\dimen10
\count1=4
\ifnum #4=0 \advance \count1 by -1 \else \fi
\ifnum #3=0 \advance \count1 by -1 \else \fi
\ifnum #2=0 \advance \count1 by -1 \else \fi
\ifnum #1=0 \advance \count1 by -1 \else \fi
\multiply \dimen8 by \count1
\raise 3mm\hbox{\vtop to \dimen8{
\hbox to \dimen1{\vsize=\dimen10\leaders\Ybox\hfil}
\hbox to \dimen2{\vsize=\dimen10\leaders\Ybox\hfil}
\hbox to \dimen3{\vsize=\dimen10\leaders\Ybox\hfil}
\hbox to \dimen4{\vsize=\dimen10\leaders\Ybox\hfil}
\vfil}}}

\def\Yngd #1 #2 #3 #4 {\Young #1 #2 #3 #4 }
\def\Yngc #1 #2 #3 {\Young #1 #2 #3 0 }
\def\Yngb #1 #2 {\Young #1 #2 0 0 }
\def\Ynga #1 {\Young #1 0 0 0 }


\dimen11=.8mm
\dimen19=.2pt
\dimen20=\dimen11 \advance \dimen20 by \dimen19
\def\ybox{\hbox to \dimen20{\boxx{\dimen11}\hfil}}
\dimen18=\dimen11 \multiply \dimen18 by 4

\def\young #1 #2 #3 #4 {
\baselineskip=\dimen11
\lineskiplimit=-10pt
\dimen1=\dimen20 \dimen2=\dimen20 \dimen3=\dimen20
\dimen4=\dimen20
\multiply \dimen1 by #1 \multiply \dimen2 by #2 \multiply \dimen3 by #3
\multiply \dimen4 by #4
\advance \dimen1 by \dimen19 \advance \dimen2 by \dimen19 \advance
\dimen3 by \dimen19 \advance \dimen4 by \dimen19
\dimen18=\dimen20
\count1=4
\ifnum #4=0 \advance \count1 by -1 \else \fi
\ifnum #3=0 \advance \count1 by -1 \else \fi
\ifnum #2=0 \advance \count1 by -1 \else \fi
\ifnum #1=0 \advance \count1 by -1 \else \fi
\multiply \dimen18 by \count1
\raise 1.2mm\hbox{\vtop to \dimen18{
\hbox to \dimen1{\vsize=\dimen11\leaders\ybox\hfil}
\hbox to \dimen2{\vsize=\dimen11\leaders\ybox\hfil}
\hbox to \dimen3{\vsize=\dimen11\leaders\ybox\hfil}
\hbox to \dimen4{\vsize=\dimen11\leaders\ybox\hfil}
\vfil}}}

\def\yngd #1 #2 #3 #4 {\young #1 #2 #3 #4 }
\def\yngc #1 #2 #3 {\young #1 #2 #3 0 }
\def\yngb #1 #2 {\young #1 #2 0 0 }
\def\ynga #1 {\young #1 0 0 0 }


\def\nboxx#1{\dimen99=#1 \dimen98=\dimen99 \advance \dimen98 by -.8pt
\hbox{\vrule\vbox to \dimen99{\hrule width \dimen98 \vfil
\hrule width \dimen98}\vrule}}

\def\Ynbox{\hbox to \dimen10{\hfil\nboxx{\dimen0}\hfil}}

\def\arbyoung{
\baselineskip=0pt
\dimen1=\dimen10 \dimen2=\dimen10 \dimen3=\dimen10
\multiply \dimen1 by 5 \multiply \dimen2 by 4 \multiply \dimen3 by 3
\advance \dimen1 by 1cm \advance \dimen2 by 1cm \advance \dimen3 by 1cm
\dimen8=\dimen10
\multiply \dimen8 by 3
\lower 6mm\hbox{\vbox to \dimen8{
\hbox to 8cm{\hbox to \dimen1{\vsize=\dimen10\Ynbox\hfil\hbox{$\cdots$}
\hfil\Ynbox}\hfil\hbox to 2cm{\vsize=\dimen10 \hfil $\nu_1$ boxes}}
\hbox to 8cm{\hbox to \dimen2{\vsize=\dimen10\Ynbox\hfil\hbox{$\cdots$}
\hfil\Ynbox}\hbox to 2cm{\vsize=\dimen10 \hfil $\nu_2$ boxes}}
\hbox to 8cm{\hbox to \dimen3{\vsize=\dimen10\Ynbox\hfil\hbox{$\cdots$}
\hfil\Ynbox}\hbox to 2cm{\vsize=\dimen10 \hfil $\nu_3$ boxes}}
\vfil}}
}

\catcode`@=11
\def\ifundefined#1{\expandafter\ifx\csname
                        \expandafter\eat\string#1\endcsname\relax}
\def\atdef#1{\expandafter\def\csname #1\endcsname}
\def\atedef#1{\expandafter\edef\csname #1\endcsname}
\def\atname#1{\csname #1\endcsname}
\def\ifempty#1{\ifx\@mp#1\@mp}
\def\ifatundef#1#2#3{\expandafter\ifx\csname#1\endcsname\relax
                                  #2\else#3\fi}
\def\eat#1{}

\newcount\refno \refno=1
\def\labref #1 #2 #3\par{\atdef{R@#2}{#1}}
\def\lstref #1 #2 #3\par{\atedef{R@#2}{\number\refno}
                              \advance\refno by1}
\def\txtref #1 #2 #3\par{\atdef{R@#2}{\number\refno
      \global\atedef{R@#2}{\number\refno}\global\advance\refno by1}}
\def\doref  #1 #2 #3\par{{\refno=0
     \vbox {\everyref \item {\reflistitem{\atname{R@#2}}}
            {\d@more#3\more\@ut\par}\par}}\vskip\refskip }
\def\d@more #1\more#2\par
   {{#1\more}\ifx#2\@ut\else\d@more#2\par\fi}
\let\more\relax
\let\everyref\relax  
\newdimen\refskip  \refskip=\parskip
\let\REF\labref
\def\@cite #1,#2\@ver
   {\eachcite{#1}\ifx#2\@ut\else,\@cite#2\@ver\fi}
\def\citeform#1{\lbrack{\bf#1}\rbrack}
\def\cite#1{\citeform{\@cite#1,\@ut\@ver}}
\def\eachcite#1{\ifatundef{R@#1}{{\tt#1??}}{\atname{R@#1}}}
\def\defonereftag#1=#2,{\atdef{R@#1}{#2}}
\def\defreftags#1, {\ifx\relax#1\relax \let\next\relax \else
           \expandafter\defonereftag#1,\let\next\defreftags\fi\next }

\newdimen\refskip  \refskip=\parskip
\def\@utfirst #1,#2\@ver
   {\author#1,\ifx#2\@ut\afteraut\else\@utsecond#2\@ver\fi}
\def\@utsecond #1,#2\@ver
   {\ifx#2\@ut\andone\author#1,\afterauts\else
      ,\author#1,\@utmore#2\@ver\fi}
\def\@utmore #1,#2\@ver
   {\ifx#2\@ut\and\author#1,\afterauts\else
      ,\author#1,\@utmore#2\@ver\fi}
\def\authors#1{\@utfirst#1,\@ut\@ver}
\def\citeform#1{{\bf\lbrack#1\rbrack}}
\let\everyref\relax            
\let\more\relax                
\let\reflistitem\citeform
\catcode`@=12
\def\Bref#1 "#2"#3\more{\authors{#1}:\ {\it #2}, #3\more}
\def\Gref#1 "#2"#3\more{\authors{#1}\ifempty{#2}\else:``#2''\fi,
                             #3\more}
\def\Jref#1 "#2"#3\more{\authors{#1}:``#2'', \Jn#3\more}
\def\inPr#1 "#2"#3\more{in: \authors{\eds#1}:``{\it #2}'', #3\more}
\def\Jn #1 @#2(#3)#4\more{{\it#1}\ {\bf#2},\ #4\ (#3)\more}
\def\author#1. #2,{#1.~#2}
\def\sameauthor#1{\leavevmode$\underline{\hbox to 25pt{}}$}
\def\and{, and}   \def\andone{ and}
\def\noinitial#1{\ignorespaces}
\let\afteraut\relax
\let\afterauts\relax
\def\etal{\def\afteraut{, et.al.}\let\afterauts\afteraut
           \let\and,}
\def\eds{\def\afteraut{(ed.)}\def\afterauts{(eds.)}}
\catcode`@=11

\newcount\eqNo \eqNo=0
\def\lasteq{\secNo.\number\eqNo}
\def\deq#1(#2){{\ifempty{#1}\global\advance\eqNo by1
       \edef\n@@{\lasteq}\else\edef\n@@{#1}\fi
       \ifempty{#2}\else\global\atedef{E@#2}{\n@@}\fi\n@@}}
\def\eq#1(#2){\edef\n@@{#1}\ifempty{#2}\else
       \ifatundef{E@#2}{\global\atedef{E@#2}{#1}}%
                       {\edef\n@@{\atname{E@#2}}}\fi
       {\rm(\n@@)}}
\def\deqno#1(#2){\eqno(\deq#1(#2))}
\def\deqal#1(#2){(\deq#1(#2))}
\def\eqback#1{{(\advance\eqNo by -#1 \lasteq)}}

\def\eqgroup(#1){{\global\advance\eqNo by1
       \edef\n@@{\lasteq}\global\atedef{E@#1}{\n@@}}}

\outer\def\iproclaim#1/#2/#3. {\vskip10pt plus5pt \par\noindent
     {\bf\dpcl#1/#2/ #3.\ }\begingroup \interlinepenalty=1000
\lessblank\it}
\newcount\pcNo  \pcNo=0
\def\lastpc{\number\pcNo} 

\def\dpcl#1/#2/{\ifempty{#1}\global\advance\pcNo by1
       \edef\n@@{\lastpc}\else\edef\n@@{#1}\fi
       \ifempty{#2}\else\global\atedef{P@#2}{\n@@}\fi\n@@}
\def\pcl#1/#2/{\edef\n@@{#1}%
       \ifempty{#2}\else
       \ifatundef{P@#2}{\global\atedef{P@#2}{#1}}%
                       {\edef\n@@{\atname{P@#2}}}\fi
       \n@@}
\def\Def#1/#2/{Definition~\pcl#1/#2/}
\def\Thm#1/#2/{Theorem~\pcl#1/#2/}
\def\Lem#1/#2/{Lemma~\pcl#1/#2/}
\def\Prp#1/#2/{Proposition~\pcl#1/#2/}
\def\Cor#1/#2/{Corollary~\pcl#1/#2/}
\def\Exa#1/#2/{Example~\pcl#1/#2/}

\font\sectfont=cmbx10 
\def\secNo{00}
\newwrite\tabcont
\immediate\openout\tabcont=\jobname.toc
\def\Beginsection#1#2{\vskip\z@ plus#1\penalty-250
  \vskip\z@ plus-#1\bigskip\vskip\parskip
  \message{#2}
                        {\noindent \bf#2}
                                        \smallskip\noindent}
\def\bgsection#1. #2\par{\Beginsection{.3\vsize}{\sectfont#1.\ #2 }%
           \write\tabcont{#1&\string\it\space#2&\the\count0\string\cr}%
            \def\secNo{#1}\eqNo=0}
\def\bgssection#1. #2\par{\Beginsection{.3\vsize}{#1.\ #2 }%
            \write\tabcont{#1&#2&\the\count0\string\cr}%
            \def\secNo{#1}\eqNo=0}
\def\Acknow#1\par{\ifx\REF\doref
     \Beginsection{.3\vsize}{\sectfont Acknowledgments}%
     \write\tabcont{&\string\it\space Acknowledgments&\the\count0\string\cr}%
#1\par
     \Beginsection{.3\vsize}{\sectfont References}%
     \write\tabcont{&\string\it\space References&\the\count0\string\cr}\fi}
\catcode`@=12
\def\class#1 #2*{{#1},}
\overfullrule=0pt

\let\REF\lstref 

\REF Aff1 Aff1 \Jref\par
\REF Aff3 Aff3 \Jref\par
\REF KT2 KT2 \Jref\par
\REF Dag Dag \Jref\par
\REF WWZ WWZ \Jref\par
\REF LonS2 LonS2 \Jref\par
\REF AAMR AAMR \Jref\par
\REF AffH AffH  \Jref\par
\REF Aff2 Aff2 \Jref\par
\REF SA SA \Jref\par
\REF AKLT2 AKLT2  \Jref\par
\REF AKLT1 AKLT1 \Jref\par
\REF FNW3 FNW3 \Jref\par
\REF FNW1 FNW1 \Jref\par
\REF AAH AAH \Jref\par
\REF Kna Kna \Jref\par
\REF BJ BJ \Jref\par
\REF KBJ KBJ \Jref\par
\REF PB PB \Jref\par
\REF NB NB \Jref\par
\REF Sol Sol \Jref\par
\REF CAHS CAHS \Jref\par
\REF Ken1 Ken1 \Jref\par
\REF WH WH \Jref\par
\REF Whi1 Whi1 \Jref\par
\REF Whi2 Whi2 \Jref\par
\REF Ken2 Ken2 \Gref\par
\REF CM1 CM1 \Jref\par
\REF SS SS \Jref\par
\REF Hol Hol \Jref\par
\REF AH AH \Gref\par
\REF LY LY \Gref\par
\REF MO  MO \Gref\par
\REF Hol2 Hol2 \Jref\par
\REF Hal2 Hal2 \Gref\par
\REF Fro Fro \Jref\par
\REF FNW2 FNW2 \Jref\par
\REF Wer1 Wer1 \Jref\par
\REF Hal Hal \Jref\par
\REF AL AL \Jref\par
\REF Maj  Maj \Jref\par
\REF MG  MG \Jref\par
\REF Kle1 Kle1 \Jref\par
\REF vdB vdB \Jref\par
\REF Kle2 Kle2 \Jref\par
\REF Cas Cas \Jref\par
\REF CM2 CM2 \Jref\par
\REF FNW5 FNW5 \Jref\par
\REF Bos1 Bos1 \Jref\par
\REF KSZ2 KSZ2 \Jref\par
\REF KLT KLT \Jref\par
\REF CCK CCK \Jref\par
\REF KK KK \Jref\par
\REF LonS1 LonS1 \Jref\par
\REF Bos2 Bos2 \Jref\par
\REF Bos3 Bos3 \Jref\par
\REF Bos4 Bos4 \Jref\par
\REF FMH1 FMH1 \Jref\par
\REF FMH2 FMH2 \Jref\par
\REF Acc Acc \Jref\par
\REF AF AF \Jref\par
\REF FNW7 FNW7 \Jref\par
\REF FNW6 FNW6 \Jref\par
\REF MS1  MS1  \Jref\par
\REF MS2  MS2  \Jref\par
\REF GW GW \Gref\par
\REF Alb1 Alb1 \Jref\par
\REF Alb2 Alb2 \Jref\par
\REF KT1 KT1 \Jref\par
\REF Mat Mat \Jref\par
\REF FF FF \Gref\par
\REF KomT KomT \Jref\par
\REF Wer2 Wer2 \Gref\par
\REF HP HP \Gref\par
\REF MP MP \Jref\par
\REF LKZ LKZ \Gref\par
\REF DNR DNR \Jref\par
\REF Che Che \Bref\par

\font\BF=cmbx10 scaled \magstep 3

\vbox to \vsize{\line{\hfill 26 October 1994}
\vskip 30pt plus30pt

\centerline{\BF The spectral gap for some spin chains}\vskip 7pt
\centerline{\BF with discrete symmetry breaking}
\vskip 20pt plus0pt
\centerline{Bruno Nachtergaele}
\centerline{Department of Physics}
\centerline{Princeton University}
\centerline{Princeton, NJ 08544-0708, USA}
\centerline{E-mail: \tt bxn@math.princeton.edu}
\vskip 20pt plus30pt

\noindent {\bf Abstract}\hfill\break
We prove that for any finite set of generalized valence bond solid
(GVBS) states of a quantum spin chain there exists a translation
invariant finite-range Hamiltonian for which this set is the set of
ground states. This result implies that there are GVBS models with
arbitrary broken discrete symmetries that are described as combinations
of lattice translations, lattice reflections, and local unitary or
anti-unitary transformations. We also show that all GVBS models
that satisfy some natural conditions have a spectral gap.
The existence of a spectral gap is obtained by applying
a simple and quite general strategy for proving
lower bounds on the spectral gap of the generator of a classical
or quantum spin dynamics. This general scheme is interesting in its own
right and therefore, although the basic idea is not new, we present it
in a system-independent setting. The results are illustrated
with an number of examples.


\vskip 10pt plus30pt
\hrule width2truein
\smallskip
{\baselineskip=12pt
\noindent
Copyright \copyright\ 1994 by the author. Faithful reproduction of
this article by any means is permitted for non-commercial purposes.\par
}}\eject

\noindent
{\bf Table of contents}
\bigskip
\halign{\hsize=16.5truecm
\parindent=0pt
\hbox to .7truecm{\hfill \bf # -\ }&\hbox to 14 truecm{#\dotfill}&\hbox to
.7truecm{\hfill #}\cr
1&\it Introduction and statement of the main results &2\cr
2&\it Lower bounds for the spectral gap --- a general strategy &8\cr
3&\it Some basic facts on pure GVBS states &15\cr
4&\it The intersection property of GVBS states &18\cr
5&\it Existence of GVBS interactions: the proof of Theorem 1 &27\cr
6&\it Existence of the spectral gap: the proof of Theorem 2 &30\cr
7&\it Examples, counterexamples, and open problems &34\cr
&\it Acknowledgments&44\cr
&\it References&44\cr
}

\bigskip

\bgsection 1. Introduction and statement of the main results

Due to recent progress made by various authors it has become clear that the
variety of behaviour found in the ground states of quantum spin models is
much larger than was expected before, even in one dimension.  In particular
there has been revived interest in models with a discrete symmetry breaking
\cite{Aff1,Aff3,KT2,Dag,{WWZ},LonS2}.  A good strategy for exploring this
variety
of phenomena has been, and still is, the study of simple exactly solvable
models in as great detail as is possible.  Therefore, various authors tried
to construct models with explicitly known ground states that exhibit some
interesting properties as, e.g., a specific kind of discrete symmetry
breaking.  E.g., in \cite{AAMR} Affleck, Arovas, Marston, and Rabson
construct spin chains with nearest neighbour interactions that have ground
states with broken charge conjugation symmetry.  These ground states are
given by means of a Generalized (or eXtended) Valence Bond Solid
constructions (GVBS, or XVBS, states).

The first question addressed in this paper is the
following. Given a finite group of symmetry transformations of a
quantum spin chain and a local observable (or a finite set of local
observables) that distinguishes ground states with broken symmetry,
can one always find a model with finite range interactions which has
the prescribed symmetries and symmetry breaking ground states? The
answer is positive: a model of the GVBS type with the desired
properties can always be constructed (see \Thm1/existence/ and the
remarks following it at the end of this introduction).

It is widely believed that if a one-dimensional quantum spin model has
a finite number of ground states (typically related to one another by
a discrete symmetry) that all have a finite correlation length (i.e.
exponential decay of correlations), then there is a spectral gap above
the ground state energy that does not vanish in the thermodynamic
limit. In some cases this is rather well understood in terms of the
two-dimensional quantum field theory that describes the long-distance
and low-energy behaviour of the spin chain \cite{AffH}. The Lorentz
invariance of the quantum field theory relates the correlation length
$\xi$
in space with the gap $\Delta$ in the spectrum which governs the
decay of correlations in (imaginary) time. The only intervening
parameter is the spin-wave velocity $v$ which plays the role of the
speed of light in the relativistic theory \cite{Aff2,SA}:
$$
\Delta = v /\xi
$$
This argument is rather heuristic at this point and cannot be given
the status of a mathematical proof. In fact, one should not expect
that a unique or a finite number of ground states with a finite
correlation length is always accompanied by a spectral gap. Certain
exceptions to this rule of thumb occur, as we show in an example
in Section 7.

To give a proof of the existence of a spectral gap in the GVBS models
with discrete symmetry breaking is the second aim of this work.
\Thm2/gap/ states that under some simple conditions any quantum spin
Hamiltonian with finite range interactions that has only a finite
number of GVBS states as its ground states, indeed possesses a spectral
gap. We also show in a counterexample (Example 2 in Section 7) that
the conditions of \Thm2/gap/ are necessary or at least that they cannot
be completely omitted; there {\it are\/} quantum spin chains with nearest
neighbour interactions and a finite number of GVBS ground states, that
do {\it not\/} have a spectral gap in the thermodynamic limit.

The first proof (in an isotropic model) of the existence of a
spectral gap was given by
Affleck, Kennedy, Lieb, and Tasaki
in \cite{AKLT2,AKLT1} in a  model with a unique
ground state (the AKLT model), and in a model with two groundstates
(the Majumdar-Ghosh model). A different proof, which applies to all
GVBS models with a unique ground state, was given by Fannes,
Nachtergaele, and Werner in \cite{FNW3,FNW1}. Apart from being more general,
this proof has the advantage of providing reasonable lower bounds on
the magnitude of the
spectral gap. Good upper bounds, both for the AKLT chain
as for the spin-1 Heisenberg antiferromagnetic chain,
are usually quite easy to obtain due
to the variational principle. For the AKLT chain such upper bounds
were obtained in \cite{AAH} and \cite{Kna}. For quite some time already,
there  is also ample numerical evidence for the Haldane gap (see
\cite{BJ,KBJ,PB,NB,Sol,CAHS,Ken1}. Knabe \cite{Kna} also
provides a general argument that, in combination with sufficiently good
numerical estimates on the gap for finite volumes, also proves the
existence of a gap in thermodynamic limit. By now, very precise
numerical estimates of the spectral gap of the AKLT chain are available
\cite{WH,SA} due to the numerical algorithm developed by White
\cite{Whi1,Whi2}. Recently Kennedy obtained upper bounds of
comparable precision \cite{Ken2} by a much simpler, variational method.
Exact excited states in GVBS models are constructed only in special
cases \cite{CM1,{SS}}.

The result (\Thm2/gap/) of this paper is very much in the spirit of
\cite{FNW1} in that it also provides, in principle, a reasonable
estimate for the gap.  It is more general because it covers the case of
multiple (a finite number of) ground states.  Part of the argument
(\Thm3/gapestimate/), however, is different from \cite{FNW1} and is an
elaboration of a discussion with H.T.  Yau.  In \Thm3/gapestimate/ we
present a general strategy for obtaining lower bounds on the spectral gap,
which, we believe, could be of wider applicability.  On a more formal level
the basic structure of this argument seems to be present in all of the
proofs of the existence of a spectral gap known to me, both for quantum
spin Hamiltonians and for the generators of stochastic time evolutions of
classical spin systems \cite{Hol,AH,LY,MO}. For a review see
\cite{Hol2}.  The theorem is formulated in
a system-independent setting which does not explicitly refer to the
one-dimensionality of the system.  It also brings the proofs of
\cite{AKLT1} and \cite{FNW1} closer together, retaining the best of
both and at the same time making them more transparent. The same
theorem can also be used to give lower bounds on the finite-volume gap in
cases where the gap vanishes in the thermodynamic limit. One then obtains
a lower bound on the rate at which the gap vanishes. In some cases the
criterion for a non-vanishing spectral gap in the thermodynamic limit that
is contained in \Thm3/gapestimate/ can be shown to be sharp. We refer the
reader to Section 2 for a discussion of this general scheme and some
related work.

The existence of a spectral gap in the generator of
a classical or quantum spin dynamics, is an important property with
direct relevance for the physical behaviour of the system.
This is true for more general models than just one-dimensional quantum
spin systems. It is, e.g., a fundamental ingredient in all theories of
the fractional quantum Hall effect. (See e.g. the pseudopotential model of
Haldane \cite{Hal2}, and the work of Fr\"ohlich and coworkers
\cite{Fro}). From the mathematical point of view there are very few
techniques available to prove the existence of a gap. Also for this
reason we chose to present the general strategy, employed in this
work, in the form of an independent theorem (Section 2), hoping that
inspiration for a proof of the spectral gap in other systems might be
drawn from it.

In this paper we will be mainly dealing with a generalization of the so-called
Valence-Bond-Solid models \cite{FNW2} that we will call GVBS models.
GVBS models are special and one cannot expect them to reveal all properties
that might be found in more general models.  However there are quite a
number of aspects in which they do provide new insight.

GVBS models are special, first of all, because their exact ground states
can be constructed in an explicit way.  In general this is not possible for
quantum spin models, not even in one dimension.  Moreover they have a very
simple structure which is quite easy to picture and essentially involves
only finite-dimensional objects.  This is closely related with another
special property that the GVBS models share: the energy is minimized {\it
locally} in their ground states.  This means that, from a certain finite
length on, the minimum energy per bond in a finite interval is the same as
the minimum energy per bond for the infinite system.  On the level of the
states this property is reflected in the fact that the finite volume ground
states coincide with the restrictions to that finite volume of the infinite
volume ground states.  One can argue that for a generic (non-GVBS)
interaction, the energy is not minimized locally.  This is due to the
non-triviality of the state extension problem for quantum spin chains
\cite{Wer1}.

Let us now look at what makes the GVBS models and their ground states
interesting objects to study.  As mentioned above, the first rigorous proof
of the properties of the Haldane phase \cite{Hal,AL} (in particular the
existence of a spectral gap) was given by Affleck, Kennedy, Lieb and Tasaki
in a particular spin-1 VBS chain \cite{AKLT1,AKLT2}, which is by now called
the AKLT-model, and it is fair to say that it served as a paradigm for many
of the subsequent studies on massive quantum spin chains.  In that paper
the authors also gave a detailed analysis of some other VBS models
\cite{Maj,MG,Kle1,vdB,Kle2,Cas,CM2}, which had been studied in the literature
before, and thus
introduced a new class of quantum spin Hamiltonians for
which exact ground states with non-trivial properties can be constructed.
Since then various other VBS models were introduced
\cite{AAH,FNW2,FNW5,Bos1,KSZ2}, including  some interesting
two-dimensional models
\cite{{KLT},{CCK},{KK},{LonS1},{Bos2},{Bos3},{Bos4}}.
A rather detailed analysis of correlation functions in a class
of GVBS chains, including inhomogeneous ones, see \cite{FMH1,FMH2}.
In \cite{FNW1}
the authors give a
definition of Generalized Valence Bond Solid states (starting from
a proposal for the construction of Quantum Markov Chains by Accardi
\cite{Acc,AF} and their analysis leads to a wide variety of VBS-type
models. In particular the Generalized VBS models and the construction
of their exact ground states does not rely on invariance under SU(2)
or SU(N) or any other symmetry group, as was the case in all previous
constructions. We will review this construction in a simplified form
below. In \cite{FNW7} it is shown that this construction generates
a weakly dense subset of the set of ergodic states.
The paper \cite{FNW1} is strongly concentrated on models with a
unique ground state.  Here our aim is to extend the basic construction
of GVBS Hamiltonians to the case where the ground state degeneracy is
arbitrary but finite, and to give a proof for the existence of a
spectral gap in that case.  This is probably the most general
situation where there is indeed a gap in one-dimensional GVBS models.
One can show that the degeneracy of the ground states for a GVBS model
(as defined in \cite{FNW1}) is either finite or grows exponentially fast
with the volume.
We exclude this case from our discussion here.

In this paper by a {\it Generalized VBS-model (GVBS-model) we mean a
one-dimensional model with a translation invariant or periodic
interaction for which there exists a non-empty finite set of ground
states that minimize the energy locally on some finite length scale
and that all the states in this set can be obtained by the generalized
VBS-construction given in\/} \cite{FNW1,FNW6}, where they are called
{\it purely generated C*-finitely correlated states}.  In this paper
we prefer to start from scratch and introduce them in a way that is as
close as possible to the traditional VBS-construction as it is known
in the literature (see \eg \cite{AKLT2}).  In fact this alternative
description was already given in \cite{FNW1}.  There exist models that
satisfy our working definition of GVBS-model --- for which the main
justification is that it defines the class of models for which our
theorems apply --- in all respects except for the fact that the number of
their zero-energy ground states is infinite and in particular contains
non-translation invariant states.  The general results presented in
this paper do not apply to such models.  In some cases they are not
expected to have a gap
\cite{FNW5}, in other cases one can still show that there is a
non-vanishing gap \cite{MS1,MS2}.

Before stating the two main theorems of this paper we
now review the GVBS-con\-struc\-tion and introduce the necessary
definitions and notations.

We label the sites in the chain by integers $i\in \Ir$ and with each
site we associate a copy of the finite-dimensional Hilbert space
$\Cx\1d$, which we denote by $\HS_i$ whenever its location in the chain
is relevant.  So, if one is to consider a chain of spin $s$ variables
one has to take $d=2s+1$. For any finite set $\Lambda\subset\Ir$,
define $\HS_\Lambda=\otimes_{i\in\Lambda}\HS_i$. Let $k\leq 1$ be an
integer and $W$ a linear map $:\Cx\1d\to\Cx\1k\otimes\Cx\1k$, and let
$\phi$ be a unit vector in $\Cx\1k\otimes\Cx\1k$. For any finite
interval $[M,N]\subset\Ir$ we can define a state for that finite piece
of the spin chain by giving its expectation values for all observables
of the form $A=A_1\otimes\cdots
\otimes A_n$ where for
all $i, M\leq i\leq N,\ A_i$ is a $d\times d$ matrix with complex
entries (e.g. a spin matrix located at the site $i$). In this paper,
states (expectation values) are usually denoted by $\om$ and scalar
products by $\langle\ \cdot\ \vert\ \cdot\ \rangle$. The expectation
value of $A$ is now defined by:
$$
\om_{[M,N],\alpha,\beta}(A)=
{\langle\phi\1{\otimes N-M+2}\vert (e_\alpha\otimes WA_M W\1*\otimes
\cdots
\otimes WA_N W\1*\otimes e_\beta) \phi\1{\otimes N-M+2}\rangle
\over{\cal N}}
\deqno(1.1)$$
where ${\cal N}$ is the normalization factor and $e_\alpha$ and
$e_\beta$ are two non-negative definite $k\times k$ matrices that play
the role of boundary conditions. It is implicitly assumed that the
expression is not identically vanishing (e.g. $W$ should be different
from zero).  It can be shown that for any choice of $e_\alpha$ and
$e_\beta$ there exists an integer $p\geq 1$ such that for all $A$ the
limit
$$
\om_{\alpha,\beta}(A)=\lim_{M\to-\infty,N\to+\infty}\om_{[pM,pN],
\alpha,\beta}(A)
$$
exists and results in a well-defined state of the infinite chain.
Typically, the state $\om_{\alpha,\beta}$ is then $p-$periodic, \ie
invariant under translations in the chain over distances that are
multiples of $p$.  We will see in Section 3 how to determine the
possible values of $p$ and also how the limit points can be described
in a simple way.  Sometimes it is convenient to regroup the chain, \ie
to partition the chain into intervals of length $p$, and to consider
it as a new chain where the elementary sites are now groups of $p$
consecutive sites in the original chain. This is also a method to
construct $p-$periodic states.  All states of the chain obtained by
the construction (1.1), possibly after carrying out a regrouping of
the chain first, will be called Generalized Valence Bond Solid states
(GVBS states). As an example one can think of the AKLT model.  There
$d=3$ and $k=2$, and $W$ identifies the space $\Cx^3$ with the
subspace of $\Cx^2\otimes\Cx^2$ (where the two $\Cx^2$ are carrying a
spin $\tover12$), which corresponds to total spin $=1$.

By an {\it interaction\/} of range $\el$ for a quantum spin chain we simply
mean a self-adjoint element $h\in (\M_d)^{\otimes \el}$.  In this paper we
will always assume that $h$ is non-negative definite, which, by itself, is
no restriction because additive constants only change the value of the
ground state energy but not the ground states of the model.  The
Hamiltonian for a finite piece of the chain, say the interval $[M,N]$, is
given by:
$$
H_{[M,N]}=\sum_{i=M}^{N-\el+1}h_i
$$
where $h_i$ is a copy of $h$ located at the sites $i,i+1,\ldots
i+\el-1$ of the chain. Here we are interested in situations where
there exists at least one state $\omega$ of the infinite chain
such that $\omega(h_i)=0$ for all $i\in\Ir$, which, as was mentioned
before, really is a rather special property. For a further discussion of
this property and some general results as well as some non-GVBS
examples where it is satisfied,
see \cite{GW}. For a given $h$, denote
the set of all zero energy states of the chain by $\Face_h$,
and let us call this set the set of ground states of the model.
It is obvious that $\Face_h$ is a {\it face\/}: if three states of
the chain $\omega,\eta_1$ and $\eta_2$, satisfy the relation
$\omega=t\eta_1+(1-t)\eta_2$ for some $t\in(0,1)$, then $\omega\in
\Face_h
\Leftrightarrow
\eta_1,\eta_2\in\Face_h$. The different ground states of the model
(in the conventional sense)
are the extreme points of $\Face_h$, and they are pure states.

It can be shown that any GVBS state as defined above is a convex
combination of a finite number of pure states that are necessarily
also GVBS states: \ie for any GVBS state $\omega$ there exist a finite
number of pure GVBS states $\omega_\alpha,\alpha=1,\ldots,n$ and real
numbers $t_\alpha>0,\sum_\alpha t_\alpha=1$, such that $\omega =
\sum_\alpha t_\alpha \omega_\alpha$. For a GVBS state $\omega$ we will
denote by $\Face_\omega$ the set of all convex combinations of the
states $\omega_\alpha$ that make up $\omega$:
$\Face_\omega=\{\eta=\sum_\alpha s_\alpha\omega_\alpha \mid s_\alpha
\geq 0,
\sum_\alpha s_\alpha=1\}$. In other words  $\Face_\omega$ is the
smallest
set of states of the chain which contains $\omega$ and has the property:
for any three states $\eta,\eta_1,\eta_2$ of the chain, if there
exists a $t\in(0,1)$ such that $\eta=t\eta_1+(1-t)\eta_2$ then $\eta\in
\Face_\omega\Leftrightarrow \eta_1$ and $\eta_2\in\Face_\omega$.

It is also known that for any finite set
of GVBS states $\omega_\alpha$, any convex combination $\omega=
\sum_\alpha
t_\alpha\omega_\alpha$ is again GVBS.
For a proof of these facts we refer to \cite{FNW1} and \cite{FNW6}.

Above we were considering states $\omega$ of the infinite chain.
They are in  one-to-one correspondence with a family of density
 matrices,
one for each finite piece of the chain. We will denote by $\rho_{[M,N]}$
the density matrix in $(\M_d)^{\otimes (N-M+1)}$ such that $\omega(A)=
\tr \rho_{[M,N]} A$ for all observables of the interval $[M,N]$,
\ie linear combinations of tensor products $A_M\otimes\cdots\otimes
A_N$. The subspace of $\Cx^{\otimes(N-M+1)}$ spanned by the eigenvectors
of $\rho_{[M,N]}$ belonging to the strictly positive eigenvalues,
will be denoted by $\G_{[M,N]}$ and will be called the local support
spaces of $\om$. Obviously, $\G_{[M,N]}=\rho_{[M,N]}\HS_{[M,N]}$.

The main body of this paper is devoted to the proof of the following
two theorems.

\iproclaim/existence/ Theorem (existence of GVBS Hamiltonians).
For any GVBS state $\omega$ there exists a finite range interaction
$h$, say of interaction length $\el$, \ie $0\leq h\in(\M_d)^{\otimes
\el}$,
such that $\Face_h=\Face_\omega$ and for all intervals $[M,N]$ such
that $N-M\geq\el$, one has $\ker H_{[M,N]}=\G_{[M,N]}$, where
$H_{[M,N]}=
\sum_{i=M}^{N-l+1} h_i$ and the $\G_{[M,N]}$ are the local support
spaces of $\om$.
\eproclaim

In particular we can take for $\omega$ a convex combination of any
finite set $\omega_1,\dots,\omega_n$ of pure GVBS states. Theorem 1.1
then
says that there exists a finite range Hamiltonian such that the set
of infinite volume ground states of the Hamiltonian exactly coincides
with the set of all convex combinations of the pure states
$\omega_1,\ldots,\omega_n$. As the set of all pure translation
invariant GVBSm states is *-weakly dense in the set of all
translation invariant states \cite{FNW7},
\Thm/existence/ implies that any possible local behaviour can be
approximated arbitrarily well by a GVBS model. In particular we can
construct GVBS models with any possible kind of discrete symmetry
breaking (see the discussion in Section 5).

\iproclaim/gap/ Theorem (existence of a spectral gap).
Let $h$ be a finite range interaction such that there exists a GVBS
state $\omega$ with the property that $\Face_\omega=\Face_h$, and such
that $\ker H_{[M,N]}=\G_{[M,N]}$ for all integers $M$ and $N$ such that
$N-M$ is large enough.
Then there exists a constant $\gamma>0$ such that for all intervals
$[M,N]\subset\Ir$  one has that the second lowest eigenvalue of
$H_{[M,N]}$ is at least $\gamma$ (the lowest eigenvalue being $0$).
Moreover for any pure state $\eta\in \Face_\omega$, and any local
 observable
$X$ such that $\eta(X)=0$, one has
$$
\lim_{M,N\to\pm\infty}\eta(X^*[H_{[M,N]},X])\geq\gamma\eta(X^*X)
$$
\eproclaim

For an explicit value of $\gamma$, i.e., a lower bound on the
gap, see Section 6.

It is important to note that the property $\Face_h=\Face_\om$ does in
general not imply that $\ker H_{[M,N]}=\G_{[M,N]}$ for all integers $M$
and $N$ such that $N-M$ is large enough. Indeed sometimes the latter
property is absent and then there might be no gap directly above the
ground state in the infinite volume model (see Section 7, Example 3).

\bgsection 2. Lower bounds for the spectral gap --- a general strategy

In this section we present in the form of a simple theorem a general
strategy to obtain lower bounds for the gap in the spectrum of the
generator of a class of spin dynamics. The basic argument, or ideas
similar to it, has been used for analyzing
irreversible time evolutions of classical spin systems
\cite{Hol,AH,LY,MO} as well as in the study of the gap above the ground
state of quantum spin Hamiltonians \cite{AKLT1,FNW1}.
\Thm3/gap/ below is essentially an elaboration of a discussion with
H.-T. Yau, who explained to the author his work with S.-L. Lu on the
spectral gap in the generator of the Glauber and Kawasaki dynamics for
Ising models.
The importance of having good estimates for the spectral gap is
obvious: In the classical case the gap determines the speed with which
the dynamics (\eg think of the Glauber dynamics for the Ising model)
drives the system toward equilibrium. For quantum systems the
essential features of the low-temperature physics are determined by
the low-lying energy spectrum, in particular the gap between the
ground state energy and the first excited state.

The general strategy, as it is described below, works only for models where
the local terms in the generator of the dynamics are minimized individually
in the reference state (see condition C2). For many quantum spin
Hamiltonians this condition is not satisfied. We might hope, however, that
once the existence of a spectral gap has been established for special
models, perturbative methods could be developed (for some first steps
in this direction see
\cite{{Alb1},{Alb2},{KT1},{Mat},{FF}}), which would enable one to show the
existence of a spectral gap
for a much wider class of models.

We have in mind the usual setup where a translation invariant model is
defined by a net of local Hamiltonians $H_\Lambda$ indexed by finite volumes
$\Lambda\subset\Ir^d$,  acting on Hilbert spaces $\HS_{\Lambda}$.
As we are interested in the behaviour of the spectrum in the
thermodynamic limit, we introduce an increasing sequence of finite
volumes $\{\Lambda_n\}_{n\in\Nl}$, such that any finite volume is
eventually contained in the $\Lambda_n$, and with the convention that
$\Lambda_0=\emptyset$. For the one-dimensional GVBS models, which are the
main object of study in this paper, the typical choice for the
$\Lambda_n$ would be an increasing sequence of intervals of the form
$[1,pn]$ for some fixed integer $p$. We will always assume that
local Hamiltonians $H_\Lambda$ are defined at least for all volumes of
the form $\Lambda_n\setminus\Lambda_m$ for all $m\leq n$.
Typically they are given
by
$$
H_\Lambda=\sum_{x, S+x\subset\Lambda} h_x
$$
where $h_x$ denotes the translate over $x\in\Ir^d$ of an self-adjoint
interaction operator $h$ acting on $\HS_S$. For a one-dimensional model
with a nearest neighbour interaction, $S$ consists of just two sites,
say 0 and 1.

For the theorem of this section only three conditions are needed.  These
conditions are stated as C1-3 below.  They do not explicitly refer to the
one-dimensionality or even translation invariance of the models.  But, as
was mentioned before, one should expect their verification for some
suitable sequence $\Lambda_n$, in particular of condition C3, to be highly
non-trivial in general.  We will formulate the conditions first and then
discuss their importance.  In the case of GVBS models conditions C1 and C2
are automatically satisfied and for GVBS models with a {\it unique\/}
ground state the proof of the C3 is contained in previous work
\cite{FNW1}.

Each of the assumptions involves some finite length
$l$. We will assume that {\bf C1-C3} hold for one and the same $l$.

\item{\bf C1}
There is a constant $d_l$ for which the local
Hamiltonians satisfy:
$$
0\leq \sum_{n=l}^N H_{\Lambda_n\setminus\Lambda_{n-l}} \leq d_l H_{\Lambda_N}
\deqno(C1)$$
Often there will be an integer $r$ --- which we can interpret as a
measure of the range of the interaction --- such that for all $l\geq r$ and $N$
large enough, there is a constant $d_l$ for which \eq(C1) holds.
For a translation invariant one-dimensional system with an
interaction of range $r$ and $\Lambda_n =[1,pn]$, one could
simply take $d_l =pl-r+1$.

\item{\bf C2}
We assume that there is a non-trivial subspace
$\G_{\Lambda_n}$ of $\HS_{\Lambda_n}$ consisting of all vectors
$\psi$ such that $H_{\Lambda_n}\psi=0$. For any finite volume
$\Lambda\subset\Lambda_N$ we denote by
$G_\Lambda$ the orthogonal projection onto the space
$\G_\Lambda\otimes\HS_{\Lambda_N\setminus\Lambda }$.
The local Hamiltonians have a non-vanishing spectral gap $\gamma_l>0$:
$$
H_{\Lambda_n\setminus\Lambda_{n-l}}
\geq \gamma_l (\idty - G_{\Lambda_n\setminus\Lambda_{n-l}})
\quad \hbox{for all } n\geq n_l
\deqno(C2)$$
where $n_l$ is some appropriate constant.
In the case of one-dimensional systems with interactions of a
finite range $r$ one could take $n_l=l+r$.

By convention we
put $G_{\Lambda_0}=G_\emptyset=\idty$, and $G_{\Lambda_{N+1}}=0$.
Note that $G_\Lambda$ and $G_{\Lambda^\prime}$
commute if either $\Lambda^\prime\subset\Lambda$
(in which case $G_{\Lambda^\prime}=G_{\Lambda^\prime}G_{\Lambda}=
 G_\Lambda G_{\Lambda^\prime}$)
or $\Lambda^\prime\cap\Lambda=\emptyset$
(in which case $G_{\Lambda^\prime}G_{\Lambda}=G_{\Lambda\cup\Lambda^\prime}$).
It follows that the operators $E_n$, $n=0,\ldots N$, defined by
$$
E_n=G_{\Lambda_n}-G_{\Lambda_{n+1}}
\deqno(En)$$
form a complete family of mutually orthogonal projections,
i.e., $\sum_{n=0}^N E_n=\idty$ and $E_n E_m=\delta_{n,m}E_n$.

The third condition is the crucial one in the present context.
We present two versions of it, C3 and C3$^\prime$.
The conditions C1-3 are sufficient for the existence of a uniform
lower bound on the spectral gaps of the local Hamiltonians,
but when C3$^\prime$ holds better explicit estimates for the spectral gap
can be obtained. In the latter case C1 should also be replaced by
C1$^\prime$ stated below.

\item{\bf C3}
We assume that there exist
$\epsilon_l <1/\sqrt{l+1}$ such that
$$
\Vert G_{\Lambda_{n+1}\setminus\Lambda_{n-l}} E_n \Vert \leq
\epsilon_l \quad \hbox{for all } n\geq n_l
\deqno(C3)$$
or equivalently
$$
E_n G_{\Lambda_{n+1}\setminus\Lambda_{n-l}} E_n
\leq \epsilon_l^2 E_n
$$

\item{\bf C3$^\prime$}
There exist constants $l_0$, $n_l$,
and $\eta_l <1/\sqrt{2}$ such that
$$
\Vert G_{\Lambda_{n+p}\setminus\Lambda_{n-l}} E_n^{(p)} \Vert \leq
\eta_l\quad\hbox{for all } p\geq 1\ , n\geq n_l\ \hbox{ and } l\geq l_0
\deqno(C3')$$
where $E_n^{(p)}=\sum_{k=n}^{n+p-1} E_k$.

\item{\bf C1$^\prime$}
There is an integer $r$ --- which we can interpret as a
measure of the range of the interaction --- such that for all $l\geq r$ and $N=
lM$ large enough, there is a constant $d$, independent of $l$,  for which the
local Hamiltonians satisfy:
$$
0\leq \sum_{m=1}^M H_{\Lambda_{lm}\setminus\Lambda_{l(m-1)}} \leq d
H_{\Lambda_N}
\deqno(C1')$$

For a translation invariant one-dimensional system with an
interaction of range $r$ one
can simply take $d=2$.

\iproclaim/gapestimate/ Theorem.
\item{i)} Assume that the conditions C1-3 are satisfied for one and the same
integer $l$.
Then, for any $N$ and any $\psi
\in \HS_{\Lambda_N}$ such that $G_{\Lambda_N}\psi =0$, i.e.,
$\psi$ is orthogonal to the space of ground states of
$H_{\Lambda_N}$, one has
$$
\langle \psi\mid H_{\Lambda_N}\psi\rangle
\geq {\gamma_{l+1} \over
d_{l+1}}(1-\epsilon_l\sqrt{l+1})^2\Vert\psi\Vert^2
$$
\item{ii)}If C1$^\prime$, C2, and C3$^\prime$ hold for $l_0$,
then for all $N=l_0 M$
$$
\langle \psi\mid H_{\Lambda_N}\psi\rangle
\geq {\gamma_{2l_0} \over d}(1-\sqrt{2}\eta_{l_0})^2\Vert\psi\Vert^2
$$
\eproclaim

The proof of this theorem is rather elementary. Of course all
essential information  is hidden in the conditions C1-3.

Condition C1 is a simple assumption on the (quasi-) local structure
of the Hamiltonians and the structure of the sequence $\Lambda_n$.
It is trivial for one-dimensional systems and $\Lambda_n$ which are
intervals increasing in a regular way.

Condition C2 restricts the applicability of the method to models where
the energy is minimized locally. It is a non-frustration
condition (see \cite{Wer1,GW} for a discussion).
For quantum spin models this is the
case for ``purely ferromagnetic'' interactions (but then, as is shown below,
C3 is not
satisfied uniformly in $N$ because of the breaking of the continuous
rotation symmetry),
and the models of the Valence-Bond-Solid type studied in
\cite{AKLT1,FNW1}. It is an interesting open problem to prove the
existence of a gap under weaker versions of C2, e.g. where one
controls the corrections to local energy minimization.

The hard work is to check Condition C3 or C3$^\prime$.
C3 plays the role of a mixing condition similar to the
Dobrushin-Shlosman condition for ergodicity. It is a well-known fact
for the conditional expectations in the Gibbs state of a
one-dimensional classical spin system with finite range interactions,
and one would expect it to be generally true also for the ground states
of quantum chains under the assumption that there is sufficient
(exponential) decay of spatial correlations.

The operators $E_n$ defined in (2) are ``conditional expectations'' in
the ground state. In a model where the energy is not minimized
locally, one could still define the $E_n$ using the local restrictions
of an infinite volume limit of the local ground states.
But then, generically, C2 cannot be expected to be satisfied. When
studying a stochastic dynamics for a classical spin model, one would
define them to be conditional expectations in the equilibrium state
(see \eg \cite{LY}).

Note that $G_{\Lambda_{n+1}\setminus\Lambda_{n-l}} E_n
= G_{\Lambda_{n+1}\setminus\Lambda_{n-l}}G_{\Lambda_n}
- G_{\Lambda_{n+1}}$.
In the case of pure GVBS states, norm bounds on this quantity
are available from \cite{FNW1}, where it is show that there exist
constants $c\geq 0$ and $0\leq \lambda <1$ such that
$$
\Vert G_{\Lambda_n} G_{\Lambda_{n+1}\setminus\Lambda_{n-l}}
- G_{\Lambda_{n+1}}\Vert
\leq c\lambda^l{1+ c \lambda^l \over 1 - c\lambda^l}
$$
In this paper the main effort of proving the existence of a spectral gap
consists in showing that \eq(C3') holds
with an $\epsilon_l< 1/\sqrt{l+1}$ {\it for all large enough\/} $n$
(Section 6).

\noindent
{\bf Proof of Theorem 2.1}\nl
{\it i)\/} From the definition of the $E_n$ \eq(En) and the assumption
that $G_{\Lambda_N}\psi=0$, it immediately follows that
$$
\psi=\sum_{n=0}^{N-1}E_n\psi
\deqno(resolutionpsi)$$
and the fact that the $E_n$ are mutually orthogonal projections
implies that
$$
\Vert\psi\Vert^2=\sum_{n=0}^{N-1}\Vert E_n\psi\Vert^2
\deqno(resolutionnormpsi)$$
Define $G_{n,l}=G_{\Lambda_{n+1}\setminus\Lambda_{n-l}}$.
Due to \eq(resolutionpsi) one has the
identity
$$
\eqalign{
\Vert E_n\psi\Vert^2&=\langle\psi\mid(\idty-G_{n,l})E_n\psi\rangle
+ \langle\psi\mid G_{n,l}E_n\psi\rangle\cr
&=\langle\psi\mid(\idty-G_{n,l})E_n\psi\rangle
+ \langle\psi\mid \sum_{m=0}^{N-1} E_m G_{n,l}E_n\psi\rangle\cr
}\deqno(Enpsi)$$
Because $G_\Lambda$ and $G_{\Lambda^\prime}$
commute if either $\Lambda^\prime\subset\Lambda$ or
$\Lambda^\prime\cap\Lambda=\emptyset$, also $E_m$ commutes with
$G_{n,l}$ if either $m\leq n-l-1$ or $m\geq n+1$.
In these cases $E_m G_{n,l}E_n= G_{n,l} E_m E_n=0$, because
the $E_n$ form an orthogonal family.
Using this observation we obtain the following estimate
from \eq(Enpsi). For any choice of constants $c_1,c_2>0$:
$$\eqalign{
&\Vert E_n\psi\Vert^2 =
\langle\psi\mid(\idty-G_{n,l})E_n\psi\rangle
+ \langle \sum_{m=n-l}^{n} E_m \psi\mid G_{n,l}E_n\psi\rangle\cr
&\quad\leq {1\over 2 c_1} \langle\psi\mid(\idty-G_{n,l})\psi\rangle +
{c_1\over 2}\langle\psi \mid E_n\psi\rangle +
{1\over 2c_2} \langle\psi\mid E_n G_{n,l} E_n\psi\rangle+
{c_2\over 2}\langle\psi \mid (\sum_{m=n-l}^n E_m)^2\psi\rangle\cr
}\deqno(Enpsi2)$$
where we have applied the inequality
$$
\vert\langle\phi_1\mid\phi_2
\rangle\vert\leq {1\over 2 c}\Vert \phi_1\Vert^2 +{c\over
2}\Vert\phi_2\Vert^2\quad ,
$$
for any $c>0$, to both terms.
The first term in the right side of inequality
\eq(Enpsi2) can be estimated with the Hamiltonian due to
condition C2 \eq(C2). To the third term we apply condition C3
\eq(C3). It then follows that
$$
(2-c_1-{\epsilon_l^2\over c_2})\Vert E_n\psi \Vert^2
-c_2 \sum_{m=n-l}^n \Vert E_m\psi\Vert^2
\leq {1\over c_1 \gamma_{l+1}}\langle\psi
\mid H_{\Lambda_{n+1}\setminus\Lambda_{n-l}} \psi \rangle
$$
We now sum over $n$, use \eq(resolutionnormpsi),
and apply condition C1 \eq(C1) to obtain
$$
(2-c_1-{\epsilon_l^2\over c_2}-(l+1)c_2)\Vert\psi\Vert^2
\leq {d_{l+1}\over c_1 \gamma_{l+1}}\langle\psi
\mid H_{\Lambda_N} \psi \rangle
$$
Finally put $c_1=1-\epsilon_l\sqrt{l+1}$ and
$c_2=\epsilon_l/\sqrt{l+1}$ and
one obtains the estimate i) stated in the theorem.

\noindent
{\it ii)\/} In order to obtain the improved estimate under condition
C3$^\prime$
\eq(C3'),
one just applies {\it i)\/} of above with $l=1$ and with a ``rescaled''
increasing  sequence of finite volumes
$\tilde\Lambda_n$, defined by
$$
\tilde\Lambda_n=\Lambda_{ln}
$$
and by using $\tilde G_{n,l}=G_{\tilde\Lambda_{n+1}\setminus
\tilde\Lambda_{n-1}}$ instead of $G_{n,l}$, and the obvious
relations $\tilde\eta_k=\epsilon_{lk}, \tilde\gamma_k=\gamma_{lk}$.
\QEDD

For the GVBS models it is easy
to show (see the proof of \Prp10// in Section 6) that a uniform lower
bound on the spectral gap of the finite-volume Hamiltonians implies
the existence of spectral gap --- bounded from below by the same lower
bound --- in the thermodynamic limit, i.e., in the spectrum of the
Hamiltonian in the GNS representation of one of the finitely many pure
infinite-volume ground states. The same relation holds for any model
with finite range interactions and for which the infinite volume ground
states can be obtained as limits of finite volume pure ground states.
The proof of \Thm/gap/ is therefore reduced to showing that the
conditions {\bf C1-3$^\prime$} hold for the GVBS models under consideration.

We conclude this section with some remarks on the quality of the lower
bounds for the gap that are obtained in \Thm/gap/. At the same time we
will illustrate with an example that one can also use these estimates
in situations where there is no gap in the thermodynamic limit. The
simplest example of this situation is the spin-1/2 Heisenberg
ferromagnetic chain. In this case \Thm/gap/ still gives a lower bound for
the finite volume gaps which is the correct order of magnitude as a
function of the size of the finite system.

The Hamiltonian of the spin-1/2 Heisenberg ferromagnetic chain of length
$N+1$, which acts on $(\Cx^2)^{\otimes (N+1)}$, can be written as follows:
$$
H_N=\sum_{i=1}^N {1\over 2}(\idty-T_{i,i+1})
$$
where $T_{i,i+1}$ is the permutation operator that interchanges the states
at sites $i$ and $i+1$.
{}From this formula for the Hamiltonian it immediately follows
that the ground state projection for an interval $[a,b]$, $G_{[a,b]}$,
is the orthogonal projection onto the space of permutation symmetric states.
For $M=1,\ldots,N$, define $\epsilon^{(N)}_1$ by
$$
\epsilon^{(M)}_1=\Vert G_{[1,M]} G_{[M,M+1]}-G_{[1,M+1]}\Vert
=\Vert  G_{[M,M+1]}G_{[1,M]}-G_{[1,M+1]}\Vert
$$
\Thm/gap/ i) and the remarks above imply that the gap
$\gamma_N$ of $H_N$ satisfies
$$
\gamma_N \geq {\gamma_2\over 2}(1-\epsilon_1\sqrt{2})
\deqno(bound)$$
where
$$
\epsilon_1=\sup_{1\leq M\leq N}\epsilon^{(M)}_1
$$
$\gamma_2$ is the gap of ${1\over 2}(\idty-T_{i,i+1})$ which is 1.
A straightforward spin-wave upper bound for the gap
for large $N$ is $\gamma_N\leq $Constant$ /N$. The bound \eq(bound)
yields non-trivial information only if $\epsilon_1 < 1/\sqrt{2}$. On the
other any $\epsilon_1 < 1/\sqrt{2}$ {\it uniform\/} in $N$ would imply
the existence of a spectral gap in the thermodynamic limit. The following
lemma shows that, in general, this critical value of $\epsilon_1=
1/\sqrt{2}$ is optimal.

\iproclaim/sharp/ Lemma.
For the spin-1/2 Heisenberg ferromagnetic chain on
an interval of length $N+1$, $N\geq 2$,  we have
$$
\epsilon_1 = {1\over\sqrt 2}\sqrt{{N^2 -1  \over N^2 +N }}
$$
\eproclaim
\proof:
Observe that $G_{[M,M+1]}G_{[1,M]}$ and $G_{[1,M+1]}$ commute
and that
$$
G_{[M,M+1]}G_{[1,M]}G_{[1,M+1]}= G_{[1,M+1]}
$$
$\epsilon^{(M)}_1$ can therefore be computed as
$$
\sup_\psi {\Vert G_{[M,M+1]}G_{[1,M]}\psi\Vert\over \vert\psi\Vert}
$$
where the $\sup$ is taken over $0\neq \psi $ such that
$G_{[1,M+1]}\psi=0$. Obviously it is sufficient to consider $\psi$
satisfying $G_{[1,M]}\psi=\psi$. Due to the SU(2) invariance of all
operators we conclude that the $\sup$ must be attained for the vector
$$
\psi=\sum_{x=1}^M S_{x,M+1}=MD_{M+1}-\sum_{x=1}^M D_x
$$
where $S_{x,y}= D_y-D_x$ and $D_z =S^-_z\mid \hbox{all up}\rangle$, for
$z=1,\ldots,M+1$.
We then just have to compute $G_{[M,M+1]}\psi$:
$$
\phi\equiv G_{[M,M+1]}\psi
={1\over 2}(M-1)(D_M + D_{M+1})-\sum_{x=1}^{M-1} D_x
$$
It is then trivial to verify that $\Vert \psi\Vert^2=M^2+M$
and $\Vert\phi\Vert^2=(M-1)^2/2 +(M-1)$.
Hence
$$
\epsilon^{(M)}_1=\sqrt{\Vert\phi\Vert^2\over \Vert\psi\Vert^2}
={1\over\sqrt{2}}\sqrt{{M^2-1\over M^2+M}}
$$
As $\epsilon^{(M)}_1$ is monotone increasing in $M$ its supremum,
$\epsilon_1$, is attained in $M=N$.
\QED

Note that the gap estimate \eq(bound) for finite volumes is of the same
order in $N$ as the upper bound from spin waves. Estimates of the
number of low energy states in the Heisenberg and other models
are given in \cite{KomT}. In Section 6 we will compare the lower bounds
on the spectral gap of GVBS models that follow from \Thm/gapestimate/
with previous work on GVBS models.

\bgsection 3. Some basic facts on pure GVBS states

Here we collect the basic properties of pure GVBS states
that we will need in the sequel. Proofs can be found in \cite{FNW1}.
For a review on GVBS states see \cite{Wer2}.

Throughout this section $\omega$ is a pure, translation invariant
state of the infinite chain. Let $\HS_i\cong\Cx^d$ denote the Hilbert
space at a site $i\in\Ir$, and for any finite subset
$\Lambda\subset\Ir$, we define
$\HS_\Lambda=\bigotimes_{i\in\Lambda}\HS_i$. In particular, for $M\leq
N\in\Ir$, $\HS_{[M,N]}$ denotes the state space of a finite piece of
the chain of length $N-M+1$. For convenience we put
$\HS_\emptyset=\Cx$, and for any Hilbert space $\HS$, we will identify
$\Cx\otimes\HS$ and $\HS\otimes\Cx$ with $\HS$ itself. Let $\M_d$ denote
the complex $d\times d$ matrices.

Suppose $\omega$ is a state obtained by the GVBS construction as
outlined in Section 1, \ie there is a $k\geq 1$, a linear map
$W:\Cx^d\to\Cx^k\otimes\Cx^k$, and a vector $\phi\in\Cx^k\otimes\Cx^k$
such that
$$\eqalignno{
&\omega(A_1\otimes\cdots\otimes A_n)=&(3.1)\cr
&\lim_{\matrix{\scriptstyle M\to -\infty\cr \scriptstyle N\to +\infty\cr}}
{\langle\phi^{\scriptscriptstyle\otimes N-M+2}\vert P_{M,N}\otimes
W\idty_d W^*\otimes\cdots WA_1 W^*\cdots WA_nW^*\cdots\otimes Q_{M,N}\mid
\phi^{\scriptscriptstyle\otimes N-M+2}\rangle\over{\cal N}(M,N)}\cr
}$$
where $0\leq P_{M,N},Q_{M,N}\in \M_k$ are chosen in one of the
possible ways to obtain a well-defined limiting state $\omega$.
It is then shown in \cite{FNW1} (Lemma 3.5 combined with Propositon 3.7),
that without loss of
generality we can assume that the following equations are satisfied:
$$
(\id_{\M_k}\otimes\Phi)(W\idty_d W^*\otimes\idty_k)=\idty_k
\deqno(3.2a)
$$
and
$$
(\Phi\otimes\id_{\M_k})(\idty_k\otimes W\idty_d W^*)=\idty_k
\deqno(3.2b)
$$
where $\Phi$ is the map $\M_k\otimes\M_k\to\Cx$ defined by
$\Phi(X)=\langle\phi\vert X\phi\rangle$ and $\id_{\M_k}$ denotes
the identity map of $\M_k$, \ie $\id_{\M_k}(X)=X$. This means that
for a given state $\omega$ we can redefine our objects, such that
${\cal N}=1$ and $P_{M,N}=Q_{M,N}=\idty_k$ and such that
moreover the limit in (3.1) becomes redundant: one can take $[M,N]
=[1,n]$ to calculate the correct expectation value of $A_1\otimes
\cdots\otimes A_n$ in the thermodynamic limit:
$$
\omega(A_1\otimes\cdots\otimes A_n)=
\langle\phi^{\otimes n+1}\vert \idty_k\otimes WA_1 W^*\otimes \cdots
WA_n W^*\otimes\idty_k\mid \phi^{\otimes n+1}\rangle
$$
It is also useful to define for all $A\in\M_d$ the operator
$\E_A:\M_k\to\M_k$ by:
$$
\E_A(B)=(\id_{\M_k}\otimes\Phi)(WAW^*\otimes B)
\deqno(3.3)$$
and a state $\rho$ of $\M_k$ by: $\rho(B)=\langle\phi\mid\idty_k\otimes
B\phi\rangle$. (3.3) then becomes
$$
\omega(A_1\otimes\cdots\otimes A_n)
=\rho(\E_{A_1}\circ\cdots\circ\E_{A_n}(\idty_k))
\deqno(3.5)$$
Instead of (3.3) we can as well write:
$$
\E_A(B)=V^*A\otimes B V
\deqno(defEA)$$
with $V:\Cx^k\to\Cx^d\otimes\Cx^k$  another isometry.
It is obvious that many choices of $k,W$ and $\phi$, even under the
restrictive conditions (3.2), will lead to the same state $\omega$.
In particular, if $\omega$ can be constructed with some $W:\Cx^d\to
\Cx^k\otimes\Cx^k$, then, by a trivial extension of the maps, we will
also have representations of $\omega$ with maps $W':\Cx^d\to\Cx^{k'}
\otimes\Cx^{k'}$ with $k'>k$. A possible way to express that the
dimension $k$ is as small as possible for a given GVBS state, is the
following. Consider the subalgebra $\B$ of $\M_k$, generated by the
elements
of the form $\E_{A_1}\circ\cdots\circ\E_{A_n}(\idty_k)$, where $n\geq1$,
and $A_1,\dots,A_n\in\M_d$. Then let $k_0$ be the smallest integer such
that $\B$ can be faithfully represented as a subalgebra of $\M_{k_0}$.
The {\it minimality condition\/} we need is $k=k_0$. Under this condition
it can  be shown that the objects that appear in the GVBS construction are
uniquely determined by the state $\omega$  up to unitary equivalence
(\cite{FNW1} , Theorem 1.3).

The states obtained by (3.1) are not necessarily pure, \ie they may
have non-trivial decompositions into other states. A very tractable
characterization of the purity of $\omega$, is given in terms of the
{\it transition operator\/} $\P\equiv\E_\idty$. $\P$ is a completely
positive transformation of $\M_k$, and \eq(3.2a) just says that
$\P(\idty_k) =\idty_k$. These two properties make $\P$ a Markov
operator (\ie $\P$ is the straightforward generalization of a Markov
operator to the non-abelian context \cite{Acc}). $\P$ governs the
ergodic properties of the state $\omega$. it turns out that under the
minimality condition stated above ($k=k_0$), $\omega$ is pure iff
$\P(X)=\lambda X$ with $\vert \lambda\vert=1$ $\Rightarrow$
$\lambda=1$ and $X$ a multiple of $\idty_k$: \ie the peripheral
spectrum of $\P$ consists only of the non-degenerate eigenvalue 1 (for
a proof of the if-part see \cite{FNW1}, Propositon 5.9; for the
only-if part see \cite{FNW6} , Theorem 1.4).  So we have a very simple
criterion that tells us exactly when $\omega$ is a pure state. For
pure GVBS states very detailed results are obtained in \cite{FNW1}. In
particular it follows that there always exists a translation invariant
finite range interaction such that $\omega$ is the unique ground state
of the corresponding model and such that the Hamiltonian has a
spectral gap above the ground state. Let us go step by step and list
the essential properties of pure GVBS states that will be used in the
following sections with the aim to extend essentially these same
properties to GVBS states that are not necessarily pure.

Let again $\rho_{[M,N]}$ be the density matrix describing the {\it
pure\/} state $\omega$ restricted to the interval $[M,N]$. Then there
are $k$ real numbers, $\rho_1>0,\ldots,\rho_k>0$, such
that the non-vanishing spectrum of $\rho_{[M,N]}$ is asymptotically
equal to $\{\rho_i\rho_j\mid 1\leq i,j\leq k\}$ in the limit $N-M\to
\infty$. The relevant property which follows from this observation
is that
$$
\inf_{M\leq N}(\spec (\rho_{[M,N]})\setminus\{0\})>0
\deqno(3.6)$$
Define
$$
m_0=\inf\{m\geq 1\mid \dim\G_{[1,m]}=k^2\}
\deqno(3.7)$$
then one can show that
$\dim\G_{[M,N]}=k^2$ for all $M,N$ such that $N-M\geq m_0-1$. For $N-M\geq
m_0-1$,
let $\psi^{M,N}_{i,j}$ denote a set of normalized eigenvectors of
$\rho_{[M,N]}$, belonging to the non-zero eigenvalues. For any local
observable $A$ we then have:
$$
\lim_{M,N\to\pm\infty}\langle\psi^{M,N}_{i,j}\vert
A\psi^{M,N}_{k,l}\rangle=\omega(A)\delta_{i,k}\delta_{j,l}
\deqno(3.8)$$
One can also show that if $\omega$ and $\eta$ are two different pure
GVBS states of the same chain, and with local support vectors
$\psi^{M,N}_{i,j}$ and $\chi^{M,N}_{i,j}$ respectively, then for any
local observable $A$:
$$
\lim_{M,N\to\pm\infty}\langle\psi^{M,N}_{i,j}\mid A\chi^{M,N}_{k,l}\rangle
=0
\deqno(3.9)$$
We will give a proof of this property in Section 4 (\Lem6/disjoint/).
The orthogonal projection onto the subspace $\G_{[M,N]}$ of $\HS_{[M,N]}$
spanned by the vectors $\psi^{M,N}_{i,j}$ will be denoted by $G_{[M,N]}$.
The spaces $\G_{[M,N]}$ satisfy a nice {\it intersection property\/}:
there exists an integer $\el_0\geq 1$ such that for all $\el\geq\el_0$
and all $M,N\in \Ir$ such that $N-M\geq \el$ one has
$$
\G_{[M,N]}=\bigcap_{k=0}^{N-M-\el+1}\HS_{[M,M+k-1]}\otimes
\G_{[M+k,M+k+\el-1]}\otimes\HS_{[M+k+\el,N]}
\deqno(3.10)$$
Here $\el_0$ can always taken to be equal to $m_0+1$, with $m_0$ defined in
\eq(3.7). In some but not all cases \eq(3.10) also holds with $\el=m_0$.
The following equivalent form of \eq(3.10) is sometimes useful: for all
$\el\geq\el_0$, all $a,b,c\in\Ir$ such that $a\leq b, b+\el\leq c$, one
has
$$
\G_{[a,c]}=\G_{[a,b+\el]}\otimes\HS_{[b+\el+1,c]}\cap\HS_{[a,b]}
\otimes\G_{[b+1,c]}
\deqno(3.11)$$
The intersection property is closely related with the existence of finite
range interactions for which the state $\omega$ is the unique ground state.
Let $p\geq 1$ be a regrouping parameter and take $\el$ such that
$p\el\geq \el_0-1+p$. Let $h\in (\M_d)^{p\el}$ a non-negative definite
observable such that $\ker h=\G_{[1,p\el]}$. Define local Hamiltonians
for the regrouped chain by:
$$
H_{[pM,pN]}=\sum_{i=M}^{N-\el+1} h_{pi}
\deqno(3.12)$$
Then
$$\ker H_{[pM,pN]}=G_{[pM,pN]}
\deqno(3.13)$$
for all $M,N$ such that $N-M\geq \el$.   This we call the
{\it ground state property\/} of $\omega$. Moreover for two such
Hamiltonians $H$ and $H'$, obtained by interactions $h$ and $h'$ of
ranges $p\el$ and $p'\el'$, there exist constants $C_1$ and $C_2$ such that
for all intervals $[M,N]$, $N-M$ large enough and compatible with the
periodicity of the Hamiltonians:
$$
C_1 H_{[M,N]}\leq H'_{[M,N]}\leq C_2 H_{[M,N]}
\deqno(3.14)$$
We will also need the following property of the ground state projections
$G_{[M,N]}$: there exists a $C>0$ and a $0\leq \lambda<1$ such that
for all $\el\geq \el_0$, $a,b,c\in\Ir , a\leq b, b+\el\leq c$, one has
$$
\Vert G_{[a,c]}-(G_{[a,b+\el]}\otimes\idty_{[b+\el+1,c]})
(\idty_{[a,b]}\otimes G_{[b+1,c]})\Vert\leq C \lambda^\el
\deqno(3.15)$$
We call this the {\it commutation property\/} of the ground state
projections. Indeed (3.15) implies that the ground state projections
for two intervals that have a large intersection, almost commute.
This property is related with the ``good factorization
property'' proved in \cite{HP} for GVBS states and similar to
the factorization property
for some classical partition functions, given in \cite{MP}.

Finally, for any choice of $h$, the Hamiltonians $H_{[M,N]}$ defined in
(3.12) have a non-vanishing spectral gap: there exists a constant
$\gamma>0$ such that for all intervals $[M,N]$:
$$
(H_{[M,N]})^2\geq \gamma H_{[M,N]}\geq 0
\deqno(3.16)$$
For GVBS models, (3.16) implies a gap of at least $\gamma$ in the
spectrum of the GNS-Hamiltonian of the infinite system.

\bgsection 4. The intersection property of GVBS states

The aim of this section is to extend the intersection property
(3.10), or equivalently (3.11), to arbitrary GVBS states, \ie dropping
the condition that they are pure. This property will be essential in
the proof of \Thm1/existence/ and \Thm2/gap/. We believe that under the
condition that  the dimension of the support
spaces of the local restrictions of the ground states is bounded (or
approximately bounded), the
intersection property actually implies the existence of a spectral
gap by itself, whether the ground states are VBS-like or not.
In the next section we will prove that for a GVBS state the
intersection property is equivalent with the existence
of a finite range interaction giving rise to \eq(3.13) (the ground
state property).

\iproclaim/intersection/ Theorem (Intersection property).
Let $\omega_1,\ldots,\omega_n$ be $n$ distinct, pure GVBS states of a
quantum spin chain. Then, the support spaces $\G_\Lambda$ of any state
$\omega$ which is a convex combination of the
$\omega_1,\ldots,\omega_n$, have the intersection property, i.e., there
exists a constant $m_0$ such that for all $l,m,r$ satisfying $l\leq 1,
m\geq m_0$, and $r\geq m$ we have
$$
\G_{[l,r]}=\G_{[l,m]}\otimes\HS_{[m+1,r]}
\cap \HS_{[l,0]}\otimes\G_{[1,r]}
$$
\eproclaim

The proof of this theorem follows from the intersection property of
pure GVBS states \eq(3.10), an orthogonality property of pure GVBS
states proved in \Lem6/disjoint/, and
\Prp7/forintersection/. \Prp7/forintersection/ itself does not involve
the GVBS nature of the states directly. It is a purely geometric
property of the support spaces.

In the considerations that follow the notion of {\it overlap\/}
between Hilbert spaces will play a crucial role.  For any two
subspaces $\HS_0$ and $\HS_1$ of a Hilbert space $\HS$, we define the
overlap as follows:
$$
\O(\HS_0,\HS_1)=\sup_{\matrix{0\neq\phi\in\HS_0\cr
                             0\neq\psi\in\HS_1\cr}}
{\vert\langle\phi\mid\psi\rangle\vert\over\Vert\phi\Vert\
\Vert\psi\Vert}
\deqno(4.1)$$
The overlap is the cosine of the angle between the subspaces.
The following properties of the overlap are elementary:
\item{i)} $0\leq\O(\HS_0,\HS_1)=\O(\HS_1,\HS_0)\leq 1$
\item{ii)} $\O(\HS_0,\HS_1)<1$ if and only if $\HS_0\cap\HS_1=\{0\}$
\item{iii)} $\O(\HS_0,\HS_1)= 0$ if $\HS_0\perp\HS_1$
\item{iv)} if $\HS\subset\HS'$, then the overlap remains unchanged if
$\HS_0$ and $\HS_1$ are now considered as subspaces of $\HS'$ rather
than
of $\HS$.
\item{v)} for any Hilbert space $\K$, consider the subspaces
$\HS_0\otimes\K$
and $\HS_1\otimes\K$ of $\HS\otimes\K$; again the overlap is
unaffected:
$\O(\HS_0\otimes\K,\HS_1\otimes\K)=\O(\HS_0,\HS_1)$.
\item{vi)} if $\HS_0\subset\HS'_0\subset\HS$ and $\HS_1\subset\HS$, then
$\O(\HS_0,\HS_1)\leq\O(\HS'_0,\HS_1)$.

Using these properties of the overlap and the specific properties of
the local support spaces $\G^\alpha_{[M,N]}$ and $\G^\beta_{[M,N]}$ of
two pure GVBS states of the same chain, say  $\omega_\alpha$ and
$\omega_\beta$, it is easy to show that for all
$\el,m,r\geq 1$:
$$
\O(\G^\alpha_{[1,\el+m]}\otimes\HS_{[\el+m+1,\el+m+r]},
\HS_{[1,\el]}\otimes\G^\beta_{[\el+1,\el+m+r]})
\leq\O(\G^\alpha_{[1,m]},\G^\beta_{[1,m]})
\deqno(4.2)$$
and from the next lemma it follows that
$$\limsup_m\O(\G^\alpha_{[1,m]},\G^\beta_{[1,m]})=0\qquad\hbox{if}\
\omega_\alpha\neq \omega_\beta
\deqno(limitepsilonm)$$

\iproclaim/disjoint/ Lemma.
Let $\omega_1$ and $\omega_2$ be two pure GVBS states of the same spin
chain with single-site Hilbert space $\HS=\Cx^d$. Denote by
$\Psi^{(M,N)}$ and $\Phi^{(M,N)}$ any pair of non-zero vectors
in the range of the local density matrices of $\omega_1$ and
$\omega_2$ respectively (i.e., vectors in the local supports of
$\omega_1$ and $\omega_2$ on the interval $[M,N]$ as defined in
\eq(3.7)). Then, if $\omega_1\neq\omega_2$,
$$
\lim_{N-M\to\infty}
{\langle \Psi^{(M,N)}\mid A \Phi^{(M,N)}\rangle\over
\Vert \Psi^{(M,N)}\Vert \, \Vert\Phi^{(M,N)}\Vert } = 0
\deqno(ortho)$$
for all local observables $A$.
\eproclaim

We believe that this lemma can be proved using general disjointness
and orthogonality properties of pure translation invariant states,
without explicit reference to GVBS states. The ``proof by
computation'' below has the advantage that it also shows how to
compute the (in general non-vanishing) inner products of finite-volume
support vectors.

\proof:
Before we start developing the argument, we collect the properties
of pure GVBS states that we will need for this proof.

For $i=1,2$ let $\omega_i$ be given in terms of an isometry
$V_i:\Cx^{k_i}\to\Cx^d\otimes \Cx^{k_i}$ (see \eq(defEA)) and a
$k_i\times k_i$ density matrix $\rho_i$, and assume that these
generating objects be minimal in the sense of \cite{FNW6} (see also
Section 3). Then, it follows from \cite{FNW6} Theorem 1.5 that the
maps defined by $$
\P_i(B)=V_i^* \idty\otimes B V_i,\quad B\in\M_{k_i}
$$ (where, as before, $\M_k$ denotes the complex $k\times k$ matrices)
have trivial peripheral spectrum. i.e., $1$ is their only eigenvalue
with modulus $=1=\Vert \P_i\Vert$, and it is non-degenerate. The
corresponding eigenvector is $\idty\in\M_k$, i.e, $$
\P_i(\idty)=\idty
\deqno(invone)$$
The map $\P_i$ leaves the state $\rho_i$ invariant in the sense that
$$
\Tr \rho_i \P_i(B)=\Tr \rho_i B
\deqno(invrho)$$
and $\rho_i$ is the unique density matrix satisfying this equation.
The $\rho_i$ are faithful states (\cite{FNW1}, Lemma 2.5). In
particular, $\rho_i$ is invertible.

The local support spaces of the state $\omega_i$ (i.e., the ranges of
the local density matrices $\rho_{[M,N]}$) are spanned by the vectors
of the form
$$
\Omega^{(n)}_i(\chi_i^L,\chi_i^R)
\equiv \underbrace{W_i\otimes\cdots W_i}_{n+1}
(\chi_i^L\otimes\underbrace{\phi_i\otimes\cdots\phi_i}_{n}\otimes\chi_i^R)
\deqno(Omega)$$
where $n=N-M$, $\chi_i^L,\chi_i^R\in\Cx^{k_i}$ are arbitrary,
$\phi_i\in\Cx^{k_i}\times\Cx^{k_i}$ is defined in terms of $\rho_i$
(it is the GNS vector of the state $\rho_i$; see the proof of
Proposition 2.7 in
\cite{FNW1}). $W_i:\Cx^{k_i}\otimes\Cx^{k_i}\to\Cx^d$ is defined in
terms of $\rho_i$ and $V_i$. The crucial relation is
$$
V_i^* A\otimes B V_i=(\id_{\M_{k_i}}\otimes \langle\phi_i\mid\,\cdot\,
\mid\phi_i
\rangle)(W_i^* A W_i) \otimes B
\deqno(Pi)$$
for all $A\in\M_d$ and $B\in\M_{k_i}$.

As the dimensions of the local support spaces are finite and
independent of the size of the interval (as long as the interval is
large enough), we can suffice with proving \eq(ortho) for the spanning
set of vectors of the form
$\Psi^{(M,N)}=\Omega_1^{(N-M)}(\chi^L,\chi^R)$ and
$\Phi^{(M,N)}=\Omega_2^{(N-M)}(\chi^L,\chi^R)$, as defined in
\eq(Omega).

For simplicity let us first consider the case $A=\idty$ and put
$n=N-M$. Then
$$\eqalign{
&\langle \Omega_1^{(n)}(\chi_1^L,\chi_1^R)\mid
\Omega_2^{(n)}(\chi_2^L,\chi_2^R)\rangle\cr
&\quad = \langle \chi_1^L\otimes\underbrace{\phi_1\otimes
\cdots\phi_1}_{n}\otimes \chi_1^R\mid
\underbrace{W_1^*W_2\otimes\cdots W_1^*W_2}_{n+1}\mid
\chi_2^L\otimes\underbrace{\phi_2\otimes\cdots\phi_2}_{n}\otimes
\chi_2^R\rangle\cr
}\deqno(ip)$$
Define a linear transformation $\P_{12}$ of $\M_{k_1,k_2}$,
the $k_1\times k_2$ matrices, by
$$
\P_{12}(B)=(\id\otimes \langle\phi_1\mid\,\cdot\,\mid\phi_2\rangle)
(W_1^* W_2\otimes B )=V_1^*\idty\otimes B V_2,\quad B\in\M_{k_1,k_2}
\deqno(P12)$$
where $V_1$ and $V_2$ are the isometries satisfying \eq(Pi).
The inner product \eq(ip) can then be written in the form
$$
\langle \chi_1^L\mid \P_{12}^n((\id\otimes\langle\chi_1^R\mid
\,\cdot\,\mid\chi_2^R\rangle)(W_1^* W_2))\mid\chi_2^L\rangle
$$

In general the matrix element between local support vectors
of a local observable $A$ is of the following form
$$
\langle \Omega_1^{(n)}(\chi_1^L,\chi_1^R)\mid
A \Omega_2^{(n)}(\chi_2^L,\chi_2^R)\rangle
= \Tr C^* \P_{12}^{n-m-l}\circ \E_A^{(l)}
\circ\P_{12}^m(B)
$$
for some $C,B\in \M_{k_1,k_2}$, where $l$ is the length
of the interval on which $A$ acts non-trivially, and $\F_A^{(l)}$
is the linear transformation of $\M_{k_1, k_2}$ defined by
$$
\F_A^{(l)}(B)=\F_{A_1}\circ\cdots\circ\F_{A_l}(B),
\quad \hbox{for }A=A_1\otimes\cdots A_l,\quad  A_1,\ldots,A_l\in\M_d
$$
and
$$
\F_A(B)=V_1^* A\otimes B V_2, \quad A\in \M_d, B\in \M_{k_1,k_2}
$$

The norms of the vectors $\Omega_i^{(n)}(\chi^L,\chi^R)$ can be
calculated in the same way:
$$
\Vert \Omega_i^{(n)}(\chi^L,\chi^R)\Vert^2
=\Tr \vert\chi^L\rangle\langle\chi^L\vert
\P_i^n((\id\otimes\langle\chi^R\mid \,\cdot\,\mid\chi^R\rangle)
(W_i^* W_i))
$$
It is straightforward to show from this relation that there are
constants $C_1,C_2>0$ such that
$$
C_1\Vert\chi^L\Vert\Vert\chi^R\Vert
\leq \Vert \Omega_i^{(n)}(\chi^L,\chi^R)\Vert
\leq C_2\Vert\chi^L\Vert\Vert\chi^R\Vert
\deqno(C1C2)$$
See \cite{FNW1} for the details.  Because of the bounds \eq(C1C2)
and the considerations above, the statement of the lemma will follow
if we show that, if $\omega_1\neq\omega_2$,
$$
\lim_{n\to\infty}\Vert \P_{12}^n\Vert =0
\deqno(limP12)$$
where $\Vert\,\cdot\,\Vert$ denotes the usual norm of linear
transformations $\P$ of $\M_{k_1,k_2}$ considered as a Banach space:
$$
\Vert\P\Vert=\sup_{0\neq B\in\M_{k_1,k_2}}{\Vert \P(B)\Vert
\over \Vert B\Vert}
$$
We will make a convenient choice for the norm on $\M_{k_1,k_2}$ below.

When $\omega_1=\omega_2$ there is always a unitary $U:\Cx^{k_2}
\to\Cx^{k_1}$ such that
$$
V_2=(\idty\otimes U^*)V_1 U
\deqno(U)$$
In particular $k_1=k_2$. This is part of Theorem 1.5 of \cite{FNW6}.
{}From \eq(U) it follows that, in the case $\omega_1=\omega_2$,
$\P_{12}(U)=U$ and hence
$\Vert\P_{12}^n\Vert\geq 1$.

We now show that in general the spectral radius of $\P_{12}$ is $\leq
1$. More specifically we show that $\Vert\P_{12}\Vert\leq 1$ if we use the norm
on $\M_{k_1,k_2}$ defined by the state $\rho_2$:
$$
\Vert B\Vert=\sqrt{\Tr \rho_2 B^*B},\quad B\in \M_{k_1,k_2}
$$
This follows from Schwarz's inequality and the properties of $\P_1$
and $\P_2$. For all $B,C\in\M_{k_1,k_2}$
$$\eqalign{
\vert\Tr C^*\P_{12}(B)\vert^2
&\quad\leq\Tr\rho_2 C^* V_1^*V_1 C\,\Tr\rho_2 V_2^*(\idty\otimes
B^*B)V_2\cr
&\quad =\Tr\rho_2 C^* C \Tr\rho_2 B^*B
}\deqno(CSineq)$$
The last equality is obtained by using \eq(invone) for the map $\P_1$
and \eq(invrho) for the map $\P_2$.
Putting $C=\P_{12}(B)$ in \eq(CSineq) yields
$$
\Vert \P_{12}(B)\Vert^4\leq\Vert\P_{12}(B)\Vert^2\Vert B\Vert^2
$$
proving that indeed $\Vert\P_{12}\Vert\leq 1$.

When the spectral radius of $\P_{12}$ is strictly less than $1$, we
have that $\Vert \P_{12}^m\Vert^{1/m} <1$ for some large enough
power $m$. In this case \eq(limP12) follows and the lemma is proved.

When $\P_{12}$ has spectral radius $=1$, we complete the proof by
showing that one necessarily has $\omega_1=\omega_2$.
In that case, $\P_{12}$ has an eigenvalue
$\lambda$ with $\vert\lambda\vert =1$, i.e., there is a  $0\neq
B\in\M_{k_1,k_2}$ such that
$$
\P_{12}(B)=V_1^* \idty(\otimes B) V_2=\lambda B
\deqno(eigen)$$
This implies $\P_2(B^*B)=B^*B$ by the following argument:
$$\eqalign{
\Tr \rho_2 V_2^* (\idty\otimes B^* B)V_2
&= \Tr \rho_2 B^*B\cr
&=\Tr\rho_2\P_{12}(B)^* \P_{12}(B)\cr
}\deqno(P2BB)$$
where for the first equality we used \eq(invrho) for $\P_2$, and
the second equality follows from \eq(eigen). \eq(P2BB) can be written
as
$$
\Tr\rho_2 V_2^*(\idty\otimes B^*)(\idty- V_1^* V_1)(\idty\otimes B)V_2
=0
$$
This is the expectation of a positive operator in the faithful state
$\rho_2$ and hence
$$
V_2^*(\idty\otimes B^*B) V_2
=\P_{12}(B)^*\P_{12}(B)=B^*B
$$
The eigenvalue $1$ of $\P_2$ is non-degenerate and therefore, by
\eq(invone),
$$
B^*B=\mu\idty\in\M_{k_2}
\deqno(BB)$$
for some $0\neq \mu\in\Cx$. By interchanging the roles of $\omega_1$
and $\omega_2$ and observing that $\P_{21}(B^*)=\overline{\lambda}
B^*$, the previous argument also shows that
$$
BB^*=\mu^\prime\idty\in\M_{k_1}
\deqno(BB*)$$
Together \eq(BB) and \eq(BB*) show that $\mu=\mu^\prime >0$ and that
$U\equiv \mu^{-1/2} B$ is unitary. In particular it follows that
$k_1=k_2$.

For the eigenvector $U$ (or $B$ for that matter) of $\P_{12}$ one has
equality in the Schwarz's inequality \eq(CSineq) with $C=B=U$ and
therefore
$$
V_1 U \rho_2^{1/2}=\mu^{\prime\prime}(\idty\otimes U)V_2\rho_2^{1/2}
$$
for some complex constant $\mu^{\prime\prime}$. As $\rho_2$ is
invertible this implies that there is a unitary, which we again denote
by $U$, which intertwines the isometries $V_1$ and $V_2$ in the
following sense:
$$
V_1 U =(\idty\otimes U) V_2
$$
It follows immediately that $\P_1$ and $\P_2$ are unitarily equivalent
and, by uniqueness of the invariant state, also that $\rho_2=U^*\rho_1
U$. It is then straightforward to check, using \eq(3.5), that
$\omega_1=\omega_2$.
\QED

Consider any state $\omega$ which is a convex combination of states
$\omega_1,\ldots,\omega_n$, i.e.,
$\omega=\sum_{\alpha=1}^nt_\alpha\omega_\alpha$, with $t_\alpha>0$ for
$\alpha=1,\ldots,n$. Then, the local support spaces $\G_\Lambda$ of
$\omega$ will be given by:
$\G_\Lambda=\bigvee_{\alpha=1}^n\G^\alpha_\Lambda$, where the
$\G^\alpha_\Lambda$ are the support spaces of the $\omega_\alpha$.
The following proposition shows that the spaces $\G_\Lambda$ inherit
the intersection property from the spaces $\G^\alpha_\Lambda$. The
only extra property of the $\G^\alpha_\Lambda$ needed to prove this is
a certain estimate on the overlap between them. In particular the
states are not assumed to be GVBS states (it is an open question
whether the intersection property of its support spaces implies that a
state is GVBS). The proof of the proposition follows from two lemmas:
\Lem8/intersectionequivalence/ which gives two equivalent formulations
of the intersection property, and \Lem9/overlapestimate/ which is an
elementary inequality for the overlap of a span of subspaces in terms
of the overlap of the subspaces.

\iproclaim/forintersection/ Proposition.
For $n\geq 2$, let $\omega_1,\dots,\omega_n$ be $n$ distinct
translation invariant states of a spin chain, whose support spaces
satisfy the intersection property
\eq(3.10) for some $m_0$, i.e., for all $l,m,r$, $l\leq1, m\geq m_0$, and
$r\geq m $:
$$
(\G^\alpha_{[\el,r]}=\G^\alpha_{[\el,m]}\otimes\HS_{[m+1,r]})\cap
(\HS_{[\el,0]}\otimes\G_{[1,r]})\deqno(4.4)
$$
Furthermore assume that for $\alpha\neq \beta$
$$
\O(\G^\alpha_{[1,m_0]},\G^\beta_{[1,m_0]})<{1\over n-1}
$$
Then there the spaces $\G_\Lambda=\bigvee_{\alpha=1}^n \G^\alpha_\Lambda$
satisfy the same intersection property \eq(4.4).
\eproclaim
\proof:
By \Lem8/intersectionequivalence/ and property ii) of the overlap we only
have to prove that for all $\alpha$
$$
\O(\K^\alpha, \bigvee_{\beta\neq \alpha} \K^\beta)<1
$$
where $\K^\alpha=(\G^\alpha_{[l,m]}\otimes\HS_{[m+1,r]})
\vee(\HS_{[l,0]}\otimes\G^\alpha_{[1,r]})$, for $\alpha=1,\ldots,n$.
As $\K^\alpha\subset\G^\alpha_{[1,m]}\subset\G^\alpha_{[1,m_0]}$, and
due to property vi) of the overlap, it is sufficient to prove
$$
\O(\G^\alpha_{[1,m_0]}, \bigvee_{\beta\neq \alpha} \G^\beta_{[1,m_0]})<1
$$
for all $\alpha=1,\ldots,n$. This follows from \Lem9/overlapestimate/
and the assumption on the mutual overlaps of the $\G^\alpha_{[1,m_0]}$
stated in the proposition.
\QED

A family of subspaces $\{\G^\alpha\}$ is called {\it independent\/} if for any
$\psi\in\bigvee_\alpha \G^\alpha$ of the form
$\psi=\sum_\alpha\psi^\alpha$ with $\psi^\alpha
\in\G^\alpha$, one has $\psi=0\Rightarrow \psi^\alpha=0$ for all
$\alpha$ or, equivalently, if the decomposition  $\psi=\sum_\alpha\psi_\alpha$
is unique. The property of independence is also equivalent with
$$
\G^\beta\cap\bigvee_{\alpha\neq\beta}\G^\alpha=\{0\}\quad\hbox{for all }
\beta
$$

The following provides us with two equivalent formulations of the
intersection property.

\iproclaim/intersectionequivalence/ Lemma.
Let $\HS_L,\HS_M$ and $\HS_R$ be Hilbert spaces and $I$ an index set,
and
let  $\{\G^\alpha_{LM}\subset\HS_L\otimes\HS_M\mid\alpha\in I\}$ and
 $\{\G^\alpha_{MR}\subset\HS_M\otimes\HS_R\mid\alpha\in I\}$ be two
families
of independent subspaces. Then the following three
properties are equivalent:

\item{i)} $(\bigvee_\alpha \G^\alpha_{LM}\otimes\HS_R)\cap
(\HS_{L}\otimes\bigvee_\alpha\G^\alpha_{MR})=\bigvee_\alpha
(\G^\alpha_{LM}\otimes\HS_R)\cap(\HS_L\otimes\G^\alpha_{MR})$

\item{ii)} $((\G^\beta\otimes\HS_R)\vee(\HS_L\otimes\G^\beta_{MR}))
\cap\bigvee_{\alpha\neq\beta}(\G^\alpha_{LM}\otimes\HS_R)\vee(\HS_L\otimes
\G^\alpha_{MR})=\{0\}$ for all $\beta\in I$.

\item{iii)} the subspaces
$(\G^\alpha_{LM}\otimes\HS_R)\vee(\HS_L\otimes
\G^\alpha_{MR})$ of $\HS_L\otimes\HS_M\otimes\HS_R$ also form an
independent family.

\eproclaim
\proof:

{\bf i) $\Rightarrow$ ii)}

Define for all $\alpha$ the space $\G^\alpha_{LMR}$ by
$$
\G^\alpha_{LMR}=
(\G^\alpha_{LM}\otimes\HS_R)\cap(\HS_L\otimes\G^\alpha_{MR})
$$
Take any $\beta\in I$ and any
$$
\psi\in(\G^\beta_{LM}\otimes\HS_R)\vee(\HS_L\otimes
\G^\beta_{MR})
\cap\bigvee_{\alpha\neq\beta}(\G^\alpha_{LM}\otimes\HS_R)\vee(\HS_L\otimes
\G^\alpha_{MR})
$$
Then there exist $\psi^\alpha_{LM}\in\G^\alpha_{LM}\otimes\HS_R$ and
$\psi^\alpha_{MR}\in\HS_L\G^\alpha_{MR}$, for all $\alpha\in I$, such that
$\psi=\psi^\beta_{LM} +\psi^\beta_{MR}=\sum_{\alpha\neq\beta}
\psi^\alpha_{LM} +\psi^\alpha_{MR}$.  Put $\xi=-\psi^\beta_{LM}+
\sum_{\alpha\neq\beta}\psi^\alpha_{LM} =\psi^\beta_{MR}-
\sum_{\alpha\neq\beta}\psi^\alpha_{MR}$. Then obviously
$$
\xi\in
\bigvee_\alpha(\G^\alpha_{LM}\otimes\HS_R)\cap
\bigvee_\alpha(\HS_L\otimes \G^\alpha_{MR})
$$
and hence, by i),
$\xi\in\bigvee_\alpha\G^\alpha_{LMR}$. From the definition of the
spaces
$\G^\alpha_{LMR}$ and the independence of either the $\G^\alpha_{LM}$
or
the $\G^\alpha_{MR}$ it follows that the spaces $\G^\alpha_{LMR}$ also form
a family of independent subspaces of $\HS_L\otimes\HS_M\otimes\HS_R$.
Using this
one immediately concludes that in the decompositions
$\xi=\sum_\alpha\xi^\alpha_{LM}=\sum_\alpha\xi^\alpha_{LMR}
=\sum_\alpha\xi^\alpha_{MR}$, with
$\xi^\alpha_{LM}\in\G^\alpha_{LM}\otimes
\HS_R$, $\xi^\alpha_{LMR}\in\G^\alpha_{LMR}$ and
$\xi^\alpha_{MR}\in\HS_L
\otimes\G^\alpha_{MR}$, one must actually have $\xi^\alpha_{LM}=
\xi^\alpha_{LMR}=\xi^\alpha_{MR}$ for all $\alpha\in I$. Comparing
the definition of $\xi$ with the employed decompositions of $\psi$
we obtain
$\psi=\psi^\beta_{LM}+\psi^\beta_{MR}=-\xi^\beta_{LM}+\xi^\beta_{MR}
=0$.

{\bf ii) $\Rightarrow$ i)}

Now take
$$
\psi\in \bigvee_\alpha(\G^\alpha_{LM}\otimes\HS_R)\cap
\bigvee_\alpha(\HS_L\otimes \G^\alpha_{MR})
$$
$\psi$ then has the
decompositions
$\psi=\sum_\alpha\psi^\alpha_{LM}=\sum_\alpha\psi^\alpha_{MR}$,
with $\psi^\alpha_{LM}\in \G^\alpha_{LM}\otimes\HS_R$ and
$\psi^\alpha_{MR}\in \HS_L\otimes\G^\alpha_{MR}$ for all $\alpha\in
I$.
For any $\beta\in I$, put $\xi=\psi^\beta_{LM}-\psi^\beta_{MR}=
\sum_{\alpha\neq\beta}\psi^\alpha_{MR}-\psi^\alpha_{LM}$. It is then
obvious
that
$$
\xi\in  (\G^\beta\otimes\HS_R)\vee(\HS_L\otimes\G^\beta_{MR})
\cap\bigvee_{\alpha\neq\beta}(\G^\alpha\otimes\HS_R)\vee(\HS_L\otimes
\G^\alpha_{MR})
$$
and by ii) this implies $\psi^\beta_{LM}=\psi^\beta_{MR}$.
As $\beta\in I$ is arbitrary, we can conclude that
$\psi\in\bigvee_\alpha
\G^\alpha_{LMR}$. So, we have shown that
$$
\bigvee_\alpha(\G^\alpha_{LM}
\otimes\HS_R)\cap\bigvee_\alpha(\HS_L\otimes \G^\alpha_{MR})\subset
\bigvee_\alpha\G^\alpha_{LMR}
$$
The opposite inclusion is trivial from the definition of the
$\G^\alpha_{LMR}$.

{\bf ii) $\Leftrightarrow$ iii)} This equivalence follows
immediately from the remark preceding the lemma.
\QED

By property ii) of the overlap independence of a family of subspaces
$\G^\alpha$ is equivalent with
$$
\O(\G^\alpha,\bigvee_{\beta\neq\alpha} \G^\beta)<1\quad\hbox{for all }
\alpha
$$
This inequality will hold when the mutual overlaps of the spaces
$\G^\alpha$ are sufficiently small, as is shown in the next lemma.

\iproclaim/overlapestimate/ Lemma.
Let $\G^1,\ldots,\G^n$ be $n$ subspaces of a Hilbert space $\HS$. Assume
$$
\O(\G^\alpha,\G^\beta)\leq \epsilon_{\alpha\beta}
$$
Then, if $\Vert B\Vert <1$
$$
\O(\G^n,\bigvee_{\alpha=1}^{n-1}\G^\alpha)\leq {\Vert a\Vert\over
\sqrt{1-\Vert B\Vert}}
\deqno(overlapineq)$$
where $a\in\Rl^{n-1}$ is the vector with components
$a_\alpha=\epsilon_{n\alpha}$, $\alpha=1,\ldots,n-1$,
 and $B$ is the $(n-1)\times(n-1)$
matrix with entries
$$
B_{\alpha\beta}=(1-\delta_{\alpha\beta})\epsilon_{\alpha\beta}
\quad \alpha,\beta=1,\ldots,n
\deqno(defB)
$$
In particular if all $\epsilon_{\alpha\beta}\leq \epsilon\leq 1/(n-1)$
we have
$$
\O(\G^n,\bigvee_{\alpha=1}^{n-1}\G^\alpha)\leq
{\epsilon\sqrt{n-1}\over\sqrt{1-\epsilon(n-2)}}\leq 1
$$
\eproclaim
\proof:
The proof is an elementary application of Schwarz's inequality.
Let $\psi_\alpha\in\G^\alpha,\alpha,1,\ldots,n$ be such that
$$
\Vert \sum_{\alpha=1}^{n-1}\psi_\alpha\Vert=1
\quad \hbox{ and }\quad\Vert\psi_n\vert =1
$$
We then have to prove that
$$
\vert\langle\psi_n\mid\sum_{\alpha=1}^{n-1}\psi_\alpha\rangle\vert
\leq {\Vert a\Vert\over \sqrt{1-\Vert B\Vert}}
$$
As $\Vert\psi_n\Vert =1$ we have
$$
\vert\langle\psi_n\mid\sum_{\alpha=1}^{n-1}\psi_\alpha\rangle\vert
\quad\leq \sum_{\alpha=1}^{n-1}\epsilon_{n\alpha}\Vert\psi_\alpha\Vert
\leq \Vert a\Vert \sqrt{\sum_{\alpha=1}^{n-1}\Vert\psi_\alpha\Vert^2}
\deqno(**)$$
Using the definition of the matrix $B$ we derive
$$\eqalign{
\left\vert \Vert\sum_{\alpha=1}^{n-1}\psi_\alpha\Vert^2
-\sum_{\alpha=1}^{n-1}\Vert\psi_\alpha\Vert^2\right\vert
&\leq \sum_{\alpha\neq\beta, 1}^{n-1}\vert\langle
\psi_\alpha\mid\psi_\beta\rangle\vert
\leq \sum_{\alpha\neq\beta, 1}^{n-1} \epsilon_{\alpha\beta}
\Vert\psi_\alpha\Vert \Vert\psi_\beta\Vert\cr
&\leq\Vert B\Vert\sum_{\alpha=1}^{n-1}\Vert\psi_\alpha\Vert^2\cr
}$$
As $\Vert \sum_{\alpha=1}^{n-1} \psi_\alpha\Vert=1$ this implies
$$
\sum_{\alpha=1}^{n-1}\Vert \psi_\alpha\Vert^2 \leq {1\over 1-\Vert B\Vert}
$$
Combined with \eq(**) this proves \eq(overlapineq).

It is obvious that if all $\epsilon_{\alpha\beta}\leq\epsilon$,
then $\Vert a\Vert\leq \sqrt{n-1}$, and because $B$ is symmetric and has
non-negative matrix elements, we also have
$$
\Vert B\Vert\leq\epsilon\Vert
(1-\delta_{\alpha\beta})_{\alpha,\beta=1}^{n-1}\Vert=\epsilon (n-2)
$$
\QED

\bgsection 5. Existence of GVBS interactions: the proof of Theorem 1

Recall that for a state $\omega$ of a quantum spin chain,
$\Face_\omega$ denotes the smallest (w*-closed) set of states of the
chain that contains $\omega$ and that satisfies: for any three states
$\eta,\eta_1$ and $\eta_2$ of the chain such that $\eta=t\eta_1+(1-t)
\eta_2$ for some $t\in(0,1)$, one has $\eta\in\Face_\omega
\Leftrightarrow \eta_1$ and $\eta_2\in\Face_\omega$.
Let $0\leq h$ be an interaction of range $\el$ and denote by $h_i$,
$i\in\Ir$, a copy of $h$ acting on the sites $i,i+1,\ldots,i+\el-1$ of
the chain. As before, we denote by $\Face_h$ the set of states $\eta$
of the chain such that $\eta(h_i)=0$, for all $i\in\Ir$.  We also use
the notation $\rho_\Lambda$ for the local density matrices of $\omega$
and the spaces $\G_\Lambda$ as defined in Section 3.

Let $h_i$ denote the translation over $i$ of a finite range interaction
$h\geq 0$. From the simple observation that
$$
\ker \left(H_{[M,N]}\equiv\sum_{i=M}^{N-\el+1} h_i \right)
= \bigcap_{i=M}^{N-\el+1}\HS_{[M,i-1]}
\otimes\ker h\otimes\HS_{[i+\el,N]}
$$ it follows that the spaces $\G_\Lambda$ have the intersection
property (3.10) iff there exists a finite range interaction $h\geq 0$
such that $\G_\Lambda=\ker H_\Lambda$ for all finite intervals
$\Lambda\subset\Ir$. For the infinite volume states we have the
following lemma.

\iproclaim/existenceinteraction/ Lemma.
Let $\omega$ be a translation invariant state of a chain such that the
local support spaces $\G_\Lambda$ of $\omega$ have the intersection
property \eq(3.10) for a certain $m_0$.
Assume in addition that there exists a
constant $\delta>0$ such that for all $M\leq N\in
\Ir$, $\rho_{[M,N]}\geq \delta $ on its support, i.e.
$$
(\rho_{[M,N]})^2\geq \delta\rho_{[M,N]}
\deqno(5.1)$$
Then there exists a finite range interaction $h\in(\M_d)^{\otimes m_0}$
such that $\Face_h=\Face_\omega$.
\eproclaim
\proof:
Let $\el_0$ be an integer such that (3.10) holds.  Define $h$ as the
orthogonal projection onto $\G^\perp_{[1,\el_0]}$. Then $h\geq0$ and
$\omega(h_i)=0$ for all $i\in\Ir$.  For any $\eta\in\Face_\omega$
there exists a $t\in(0,1)$ such that we can find a state $\eta'$ such
that $\omega=t\eta+(1-t)\eta'$.  It follows that $\eta(h_i)=0$ for
all $\eta\in\Face_\omega$. Hence $\Face_\omega\subset\Face_h$.

In order to
prove the opposite inclusion, take $\eta\in\Face_h$. Then, by the
intersection property, we must have that for any finite volume
$\Lambda$ the restriction $\eta_\Lambda$ of $\eta$ is a substate of
$\omega_\Lambda$, \ie there exists a constant $C_\Lambda(\eta)>0$
such that
$$
\eta_\Lambda\leq C_\Lambda(\eta)\omega_\Lambda
\deqno(5.2)$$
The condition (5.1) implies that $C_\Lambda(\eta)$ can always be taken
to $\delta^{-1}$, \ie independent of $\Lambda$ and $\eta$. It follows
that there exists a $t_\Lambda\in[\delta, 1]$, and a state $\eta'_\Lambda$
such that
$$
\omega_\Lambda=t_\Lambda\eta_\Lambda+(1-t_\Lambda)\eta'_\Lambda
\deqno(5.3)$$
Choose a sequence of intervals $\Lambda_i$, increasing to $\Ir$, such
that $\lim_{i\to\infty} t_{\Lambda_i}$ exists and equals say $t$.
Then $\omega=t\eta+(1-t)\eta'$, where $\eta'=\lim_i \eta'_{\Lambda_i}$
is well-defined because of (5.3).  As $t\geq\delta>0$ we can conclude
that $\eta\in\Face_\omega$.
\QED

We now can complete the proof of Theorem 1.1.

\noindent
{\bf proof of Theorem 1.1:}\hfill\break It follows from the
decomposition theory of GVBS states that any GVBS state can be
decomposed into a finite number of ergodic components, which are again
GVBS states.  So, if $\omega$ is a GVBS state, there are ergodic GVBS
states $\omega_1,\ldots,\omega_k$ and convex combination coefficients
$t_1,\ldots,t_k$, such that $\omega=\sum_{i=1}^k
t_i\omega_i$. Furthermore any of these ergodic GVBS states $\omega_i$
has a decomposition into $p_i$ periodic states $\omega_{i,q}$,
$q=1,\ldots,p_i$, with equal weights:
$\omega_i=\sum_{q=1}^{p_i}\omega_{i,q}$.  The states $\omega_{i,q}$
are invariant under translation over $p_i$ lattice spacings:
$\omega_{i,q}
\circ\tau_{p_i}=\omega_{i,q}$, and one also has that $\omega_{i,q}\circ
\tau_1
=\omega_{i,q+1(\mod p_i)}$.  All the $\omega_{i,q}$ are pure GVBS
states. For a proof of these properties see \cite{FNW1}, or for a
more complete account see \cite{FNW6} .

\noindent
Let $p$ be the least common multiple of $p_1,\ldots,p_k$. Then all states
$\omega_{i,q}$ are $\tau_p$-invariant.  So, consider a regrouped chain
where the sites correspond to intervals of length $p$ of the original chain.
Now we are in a situation where \Thm4// applies
with $n=\sum_{i=1}^k p_i$.
So, at the level of the regrouped chain we have the intersection property
for the ground state spaces $\G_\Lambda$ of $\omega$. By \Lem9//
this implies that there exists finite range interaction $h^{(p)}\in
(\M_d^{\otimes p})^{\otimes m_0}$, for some constant $m_0$, such that
$\Face_\omega =\{
\eta\mid \eta(h^{(p)}_{pi}) \hbox{\rm  for all } i\in\Ir\}$.
The condition \eq(5.1) in Lemma 5.1 is satisfied for GVBS states because
of \eq(3.6). Here $h^{(p)}_{pi}$
acts on the sites $pi,pi+1,\ldots,pi+p m_0+1$ of the original chain. This
does not immediately yield a translation invariant Hamiltonian for which the
states $\omega_{i,q}$ are the ground states. But the interaction can be made
translation invariant by defining
$$
h_i=\sum_{q=0}^{p-1}
h^{(p)}_{i+q} \ \in \M_d^{\otimes p m_0+p-1}
$$
Observe that due to the translation invariance of $\omega$ we have:
$$
\omega(h_i)=\sum_{q=0}^{p-1}(\omega\circ\tau_q)(h^{(p)}_i)=0
$$
Hence also $\omega_{i,q}(h)=0$ for all $i=1,\ldots,k$ and $q=1,\ldots,p_i$.
So we conclude that $\Face_\omega\subset\Face_h$. As $h\geq h^{(p)}$
we certainly have $\Face_h\subset\{\eta\mid \eta(h^{(p)}_{pi})=0\}=
\Face_\omega$
hence $\Face_h=\Face_\omega$.
\QEDD

{}From the arguments in the proof it is also clear that the interaction $h$
can be  chosen such that the Hamiltonian is invariant under all symmetries
of the set of states $\{\omega_i\mid i=,1\ldots,k\}$. This holds equally well
for broken as for unbroken symmetries. By a broken symmetry we mean a
symmetry
transformation that does not leave invariant at least one of the pure
components of the states $\omega_i$,  but such that
states  in the set $\Face_\omega$ are transformed into states of
$\Face_\omega$.
An interesting consequence of this observation is that one can construct
GVBS models with any kind of prescribed discrete symmetry, and such that
this symmetry is spontaneously broken  in the ground states of the model.
The only requirement is that there exist translation invariant (or
periodic) ergodic states of the chain (not necessarily GVBS states) which
break the symmetry, and which belong to a finite orbit of the symmetry
group, i.e., by application of the symmetry transformations one generates
a finite number of different states. The only requirement is that the
symmetries preserve the GVBS nature of the states, i.e., they should transform
GVBS
states into GVBS states. This is known to be the case for
lattice translations, lattice reflections, and local symmetries described by
unitary or anti-unitary transformations, which
includes the following examples:
\item{1)} symmetries described by a finite group of unitaries  $U\in\M_d$,
acting on the observables as $A\mapsto (U^*)^{\otimes N} A U^{\otimes N}$,
for $A$ an observable that lives on an interval of length $N$.
\item{2)} the translation symmetry of the chain
\item{3)} the reflection symmetry $R$ of the chain given by
$\Ir\ni i\mapsto -i$ and its natural lifting to the algebra of observables. A
necessary and sufficient condition for a pure GVBS state $\omega$ to be
$R$-invariant
is the following. Let $\omega$ be defined in terms of an isometry
$V:\Cx^k\to
\Cx^d\otimes\Cx^k$ by the formulae:
$$
\E_A(B)=V^* A\otimes BV\ , \ \hbox{ for all} \ A\in\M_d,\, B\in\M_k
$$
and  for $A_1,\ldots,A_i\in \M_d$
$$
\omega(A_1\otimes\cdots\otimes A_n)=\rho(\E_{A_1}\circ\cdots
\circ\E_{A_n}(\idty_k))
$$
where $\rho$ is a state of $\M_k$ satisfying $\rho(\E_\idty(B))=
\rho(B)$, and
 where
$k$ is the smallest integer for which this is possible (see the
minimality condition mentioned in Section 2).  Then $\omega$ is
$R-$invariant iff there exist orthonormal bases $\{ f_a\}$ and
$\{ e_i\}$ of $\Cx^d$ and $\Cx^k$ respectively, and
a
unitary $U\in\M_k$,
such that for all $a=,1,\dots,d,\, i,j=1,\ldots,k$, one has:
$$
\langle f_a\otimes Ue_i\mid VUe_j\rangle=\langle f_a\otimes e_j\mid
Ve_i\rangle
$$
If one fixes the vector $f_a$,the action of $V$ is given by a $k\times k$
matrix. The
above relation says that there exists a single unitary $U$ which
transforms these
$d$ matrices of dimensions $k\times k$  into their transposes.
This characterization is  an application of
the results in \cite{FNW6}.

\item{4)} symmetries described by anti-unitaries as the charge conjugation
and the chiral symmetry (cfr. Section 7 for an example)

Moreover any of the above symmetries can be considered after regrouping
the chain first and, of course, one can also form products of
the elementary symmetries described in 1-4).

\bgsection 6. Existence of the spectral gap: the proof of Theorem 2

In order to prove the existence of a spectral gap for the GVBS models
obtained in the previous section, we need to develop the arguments in
the proof of Proposition 4.1 a little bit further. This is
accomplished in Lemma 6.1, which, together with the results of
\cite{FNW1}, proves that condition {\bf C3$^\prime$} of Section 2 is
satisfied. Conditions {\bf C1} and {\bf C2} are trivially satisfied in
the situation at hand. \Thm/gap/ is then a direct consequence of
\Thm/gapestimate/, the properties of pure GVBS states proved in
\cite{FNW1}, and a simple argument to pass to the thermodynamic
limit.

For convenience we define
$$
A^\alpha_m=\sup_{\el\leq1,r\geq m}\Vert G^\alpha_{[\el,r]}-
(G^\alpha_{[\el ,m]}\otimes\idty_{[m+1,r]})(\idty_{[\el,0]}\otimes
G^\alpha_{[1,r]})\Vert
$$
$A^\alpha_m$ is the best constant in the commutation property
\eq(3.15) for the ground state projections of pure state
$\omega_\alpha$ on arbitrary finite intervals that overlap on $m$ sites.
Combination of the results in \cite{FNW1} and in \cite{FNW6} proves
that the $A^\alpha_m$ satisfy a bound of the form
$$
A^\alpha_m\leq c\lambda^m{1+ c \lambda^m \over 1 - c\lambda^m}
\deqno(boundAm)$$
for some constants $C>0$ and $0<\lambda<1$ which depend on the
state $\omega_\alpha$. $C$ can be taken to be equal to $k^2$ where
$k$ is the dimension of the auxiliary space used to define the pure
GVBS state $\omega_\alpha$ (see (3.1)). $\lambda$ can be any number
satisfying $\lambda_i < \lambda <1$ for all eigenvalues $\lambda_i\neq 1$
of the transfer operator $\P$ for the state $\omega_\alpha$.

We now first derive the commutation property for the GVBS
state $\omega$ that we need for the proof of  \Thm/gap/.
Define for all $m\geq 1$ the operator $X_m$ by
$$
X_m=\sum_{\alpha=1}^n G^\alpha_{[1,m]}
\deqno(defXm)
$$
For all $m$ such that
$$
\epsilon_{\alpha\beta}(m)\equiv\O(\G^\alpha_{[1,m]},\G^\beta_{[1,m]})
<{1\over n-1}
\deqno(epsilon)$$
we define
$$
\delta_m={\Vert \epsilon\Vert\over 1-\Vert
\epsilon\Vert}
\deqno(deltam)$$
where $\epsilon=(\epsilon_{\alpha\beta})$ is the $n\times n$ matrix
of the mutual overlaps of the spaces $\G^\alpha_{[1,m]}$ and with
$0$'s on the diagonal, i.e., we put $\epsilon_{\alpha\alpha}=0$ by
definition. Note that if $\epsilon_{\alpha\beta}\leq \epsilon$ for all
$\alpha$, one has the simple bound $\Vert\epsilon\Vert\leq (n-1)\epsilon$,
and $\delta_m\leq (n-1)\epsilon/(1-(n-1)\epsilon)$. From the properties of the
overlap it immediately follows that the $\epsilon_{\alpha\beta}$, and hence
also
$\delta_m$, are decreasing functions of $m$.

As before, $G_\Lambda$ denotes the orthogonal projection on to the
space $\bigvee_{\alpha=1}^n\G^\alpha_\Lambda$.

\iproclaim/commutation/ Lemma.
Let $m_0$ be such that
$$
\O(\G^\alpha_{[1,m]},\G^\beta_{[1,m]})
<{1\over 2(n-1)}\quad \hbox{for all }\alpha\neq\beta
\deqno(5a)$$
and let $\delta_m$ be defined as in \eq(deltam) and let $X_m$ be the
operators defined in \eq(defXm). Then

\item{i)} for all $m\geq m_0$ we have the bound
$$
 \Vert X_m-G_{[1,m]}\Vert\leq {\delta_m\over 1-\delta_m}
\deqno(boundXm)$$
\item{ii)} For all  $m\geq m_0, \el \leq 1, r\geq m$ one has
$$
\Vert G_{[\el,r]}-(G_{[\el ,m]}\otimes\idty_{[m+1,r]})
(\idty_{[\el,0]}\otimes
G_{[1,r]})\Vert\leq 4{\delta_m\over (1-\delta_m)^2}
+\sum_{\alpha=1}^n A^\alpha_m
$$
\eproclaim
\proof:
We will use the following bound for the proof of both i) and ii). Let
$\phi$ and $\psi$ be two vectors that are of the form $\phi
=\sum_{\alpha=1}^n \phi_\alpha$ and $\psi =\sum_{\alpha=1}^n \psi_\alpha$,
with $\phi_\alpha\in\G^\alpha_{\Lambda_1}$ and
$\psi_\alpha\in\G^\alpha_{\Lambda_2}$, where $\Lambda_1$ and $\Lambda_2$
are two finite volumes containing the interval $[1,m]$.
We will show that
$$
\sum_{\alpha\neq\beta}\vert\langle\phi_\alpha\mid\psi_\beta\rangle\vert
\leq \delta_m \Vert\phi\Vert\,\Vert\psi\Vert
\deqno(5d)$$
As in the proof of \Lem/overlapestimate/ we use Schwarz's inequality
to obtain
$$\eqalign{
\sum_{\alpha\neq\beta}\vert\langle\phi_\alpha\mid\psi_\beta\rangle\vert
&\leq \sum_{\alpha\neq\beta}\epsilon_{\alpha\beta}\Vert\phi_\alpha\Vert\,
\Vert\psi_\beta\Vert\cr
&\leq \Vert\epsilon\Vert\sqrt{\sum_{\alpha=1}^n\Vert\phi_\alpha\Vert^2
\sum_{\alpha=1}^n\Vert\psi_\alpha\Vert^2}\cr
}\deqno(5b)$$
Applying this inequality for $\phi_\alpha=\psi_\alpha$ one obtains
$$
\left\vert\Vert\sum_{\alpha=1}^n\phi_\alpha\Vert^2-
\sum_{\alpha=1}^n\Vert\phi_\alpha\Vert^2\right\vert
\leq
\Vert\epsilon\Vert \sum_{\alpha=1}^n\Vert\phi_\alpha\Vert^2
$$
and hence
$$
\sum_{\alpha=1}^n\Vert\phi_\alpha\Vert^2\leq {1\over
1-\Vert\epsilon\Vert}\Vert\phi\Vert^2
\deqno(5c)$$
Combining \eq(5b) and \eq(5c) we obtain \eq(5d).

\noindent
{\it proof of i):\/}
It is obvious that $X_m(\idty_{[1,m]}-G_{[1,m]})=0$ and therefore
$$
\Vert X_m-G_{[1,m]}\Vert=\sup_{\matrix{\Vert\phi\Vert=1\cr
G_{[1,m]}\phi=\phi\cr}} \Vert (X_m-G_{[1,m]})\phi\Vert
$$
Then, for $G_{[1,m]}\phi=\phi$, we can write $\phi=
\sum_\alpha\phi_\alpha$,
where $\phi_\alpha\in\G^\alpha_{[1,m]}$, for $\alpha=1,\ldots,n$ and
we have $\Vert G^\beta_{[1,m]}\phi_\alpha\Vert\leq\epsilon_m
\Vert\phi_\alpha\Vert$ if $\alpha\neq\beta$.
So, we can estimate $\Vert (X_m-G_{[1,m]})\phi\Vert=
\Vert X_m\phi-\phi\Vert$ by estimating a quantity
of the form \eq(5d):
$$\eqalign{
\Vert X_m\phi-\phi\Vert&=\Vert\sum_\alpha \phi_\alpha-\phi +\sum_{\alpha
\neq \beta}G^\beta_{[1,m]}\phi_\alpha\Vert\cr
&=\Vert\sum_{\alpha\neq \beta}G^\beta_{[1,m]}\phi_\alpha\Vert\cr
&=\sup_{\Vert\psi\Vert=1}\vert\sum_{\alpha\neq\beta}\langle\psi
\mid G^\beta_{[1,m]}\phi_\alpha\rangle\vert
\leq \sup_{\Vert\psi\Vert =1}\sum_{\alpha\neq\beta}
\vert\langle\psi_\beta\mid\phi_\alpha\rangle\vert\cr
}$$
where $\psi_\beta=G^\beta_{[1,m]}\psi$. By \eq(5d) this implies
$$
\Vert X_m-G_{[1,m]}\Vert\leq\delta_m \Vert X_m\Vert
$$
As $\Vert G_{[1,m]}\Vert=1$, this implies i).

\noindent
{\it proof of ii):\/}
By the triangle inequality it is sufficient to estimate the following
sum of three terms:
$$\eqalign{
&\quad \Vert G_{[\el,r]}-X_{r-\el+1}\Vert +\Vert X_{r-\el+1} -
\sum_\alpha(G^\alpha_{[\el ,m]}\otimes\idty_{[m+1,r]})(\idty_{[\el,0]}
\otimes G^\alpha_{[1,r]})\Vert\cr
&\qquad  +\Vert\sum_\alpha(G^\alpha_{[\el ,m]}\otimes\idty_{[m+1,r]})
(\idty_{[\el,0]}\otimes G^\alpha_{[1,r]})-
(G_{[\el ,m]}\otimes\idty_{[m+1,r]})(\idty_{[\el,0]}\otimes
G_{[1,r]})\Vert\cr
}$$
The first term is bounded above by
$\Vert G_{[l,r]}-X_{r-l+1}\Vert$ which we can estimate using i).
The second term is estimated by $\sum_{\alpha=1}^n A^\alpha_m $.
The third term can be treated as follows:
$$\eqalign{
&\Vert\sum_\alpha(G^\alpha_{[\el ,m]}\otimes\idty_{[m+1,r]})
(\idty_{[\el,0]}\otimes G^\alpha_{[1,r]})-
(G_{[\el ,m]}\otimes\idty_{[m+1,r]})(\idty_{[\el,0]}\otimes
G_{[1,r]})\Vert\cr
&\quad\leq \Vert(G_{[\el,m]}\otimes\idty_{[m+1,r]})(\idty_{[\el,0]}\otimes
G_{[1,r]}) - (X_{m-\el+1}\otimes\idty_{[m-1,r]})(\idty_{[\el,0]}\otimes
X_r)\Vert\cr
&\qquad\quad + \Vert\sum_{\alpha\neq\beta}(G^\alpha_{[\el ,m]}\otimes
\idty_{[m+1,r]})
(\idty_{[\el,0]}\otimes G^\beta_{[1,r]})\Vert\cr
}$$
By adding and subtracting
$(G_{[\el,m]}\otimes\idty_{[1,r]})(\idty_{[\el,0]}\otimes X_r)$, the triangle
inequality, and repeated use of i) we find that the first term in the
right side of the inequality above is bounded by
$$
{\delta_{r-l+1}\over 1- \delta_{r-l+1}}
+ {\delta_{r-l+1}\over (1- \delta_{r-l+1})^2}
$$
{}From  \eq(5d) it follows that the second term  is bounded
by $\delta_m/(1-\delta_m)^2$. Collecting these estimates we obtain
$$\eqalign{
&\Vert G_{[\el,r]}-(G_{[\el ,m]}\otimes\idty_{[m+1,r]})
(\idty_{[\el,0]}\otimes G_{[1,r]})\Vert\cr
&\quad \leq
{2\delta_{r-l+1}\over 1- \delta_{r-l+1}}
+ {\delta_{r-l+1}\over (1- \delta_{r-l+1})^2} +{\delta_m\over (1-\delta_m)^2}
\sum_\alpha A^\alpha_m\cr
}$$
Because of the monotonicity of $\delta_m$ this implies ii).
\QED

We now prove the existence of a non-vanishing uniform lower bound
for the spectral gap of the finite volume Hamiltonians $H_{[M,N]}$
defined by $H_{[M,N]}=\sum_{i=M}^{N-\el_0+1} h_i$, where $h$ is an
interaction of range $\el_0$ with the properties stated in Theorem 1.1.
In fact the proof of this proposition does not rely anymore
on the explicit GVBS structure of the ground states.
The gap property is a direct  consequence
of the intersection property of the local support spaces $\G_\Lambda$
and the commutation property  of the projections $G_\Lambda$ obtained
in Lemma 5.1. It is an interesting open question whether or not these
properties by themselves imply that the state is a GVBS state.

\iproclaim/prooftheorem2/ Proposition (proof of \Thm/gap/).
Under the assumptions of \Thm/gap/ there exists a constant $\gamma>0$
such that for $N-M$ large enough, the gap between the lowest and
the second lowest eigenvalue of $H_{[M,N]}$ exceeds $\gamma$.
$\gamma$ is also a lower bound for the gap of the infinite
system in any of its pure ground states.
For $m$ large enough one has the following non-trivial lower
bounds for $\gamma$:
$$
\gamma\geq {\gamma_{2m}\over 2}(1-\eta_m\sqrt{2})^2
\deqno(boundgamma)$$
where $\gamma_{2m}$ is the gap of the finite-volume Hamiltonian
$H_{[1,2m]}$ and $\eta_m$ satisfies
$$
\eta_m\leq {4(n-1)\epsilon_m\over 1-2(n-1)\epsilon_m}
+\sum_{\alpha=1}^n A^\alpha_m
\deqno(boundeta)$$
where
$\epsilon_m=\max_{\alpha\neq\beta}\O(\G^\alpha_{[1,m]},\G^\beta_{[1,m]})$,
and the $A^\alpha_m$ satisfy  the bound \eq(boundAm).
\eproclaim
How fast the first term in \eq(boundeta) vanishes as $m\to\infty$
depends on how different the states $\omega_\alpha$ are on
intervals of $m$ sites. The $A^\alpha_m$ also tend to $0$ as $m\to
\infty$, and has the same exponential behaviour as the truncated
two-point correlation functions of the states $\omega_\alpha$.
One should indeed expect two contributions of this kind
to the commutation estimate ii) of \Lem/commutation/. The
support projections $G_\Lambda$ of $\omega$ cannot be expected to
have better commutation properties than the projections of the
components $\omega_\alpha$. On the other hand, even if the
$\omega_\alpha$ have perfect commutation properties, the convex
combination $\omega$ could fail to have these properties when the
$\omega_\alpha$ are to close to one another. This happens, e.g.,
when there is breaking of a continuous symmetry as in the Heisenberg
ferromagnet.

In the case $n=1$, i.e., GVBS models with a unique ground state,
the first term in the right side of \eq(boundeta) vanishes for all $m$.
The estimate
on the infinite volume gap implied by \eq(boundgamma) and
\eq(boundeta), is then a little bit better than the one previously
obtained in \cite{FNW1}, which is with $(1-\sqrt{2}\eta_m)^2$ replaced by
$(1-2\eta_m)$. On the other hand, our estimate
\eq(boundgamma) suffers from the same overall factor $1/2$ that was
also present in \cite{FNW1}. One therefore should expect the bounds to
underestimate the infinite-volume gap by a factor $1/2$ at best.

\proof:
{}From the previous results it is straightforward to check that
the conditions {\bf C1-C3$^\prime$} (see Section 2) are satisfied
with the constants $d=2$ and $\gamma_{2m}$ as stated in the proposition.
The uniform lower bound for the gap of the finite-volume Hamiltonians
then follows directly from \Thm/gapestimate/.

In order to complete the proof we still have to show that the
finite-volume estimate  also applies to the gap of the GNS
Hamiltonian of the infinite system in one of the ground states.
This implication is rather trivial in the case at hand because any
of the pure infinite volume ground states can be obtained as
a limit of pure finite volume ground states of the Hamiltonians $H_\Lambda$
of which we proved that they have a uniformly bounded gap.
Indeed, as the range of the interaction is finite,
we have for any strictly local observable $X$ that is supported on the
volume $\Lambda_0$
$$
\eqalign{
{\lim_{\Lambda \nearrow \Ir}\omega_\alpha (X^*[H_\Lambda,X])\over
\omega_\alpha(X^* X)-\vert\omega_\alpha(X)\vert^2}&={\omega_\alpha
(X^*[H_{\Lambda_0},X])\over
\omega_\alpha(X^* X)-\vert\omega_\alpha(X)\vert^2}\cr
&={\lim_{\Lambda_1\nearrow \Ir}\omega_{\alpha,\Lambda_1}
(X^*[H_{\Lambda_1},X])
\over
\lim_{\Lambda_1\nearrow \Ir}(\omega_{\alpha,\Lambda_1}(X^* X)-\vert
\omega_{\alpha,\Lambda_1}(X)\vert^2)}\cr
&\geq \gamma\cr
}$$
Here the states $\omega_{\alpha,\Lambda_1}$ are of the form
$\langle\psi_{\alpha,\Lambda_1}\mid \,\cdot\,\mid
\psi_{\alpha,\Lambda_1}\rangle$ for some zero
eigenvector $\psi_{\alpha,\Lambda_1}$ of $H_{\Lambda_1}$.
\QED

\def\irr{irreducible}
\def\rep{representation}
\def\irrep{\irr\ \rep}
\def\cc{charge conjugation}

\bgsection 7. Examples, counterexamples, and open problems

In this section we want to show how the general results of this paper,
in particular Theorem 1.2, can be applied to a great variety of 1-D spin
Hamiltonians. Although checking the conditions of Theorem 1.2 seems very
simple, there is a subtle point that easily could be overlooked. Suppose
one has a model defined in terms of a finite-range interaction $h\geq 0$
and such that the infinite-volume zero-energy ground states of the model
are all convex combinations of a finite number of GVBS-states. The subtlety
is that this does {\it not\/} imply that $\ker H_{[M,N]}=\G_{[M,N]}$.
In other words, the finite-volume Hamiltonians might have ground states
that are not found back in the local support spaces of the infinite
volume ground states. The thermodynamic limits of these additional ground
do not go beyond the GVBS ground states we already had, but for a finite volume
they are different. An example of this situation is given in Example 2.
There we show that not only does Theorem 1.2 not apply, but that moreover
there is no gap above the ground state.

We now briefly discuss five models or families of models,
examples and
counterexamples, and also indicate some open problems.
It must be clear that our only aim is illustration and that what is
given below definitely does not exhaust the possible applications
of the theorems. We do not discuss any new examples of GVBS models
with a unique ground state, because these are completely covered by
the results in \cite{FNW1}. A recent additions to the family of GVBS models
with unique ground states is e.g. \cite{LKZ}.

We start with the well-known Majumdar-Ghosh model. Although it is a
special case of the generalized Majumdar-Ghosh model discussed in example 1bis,
we prefer to discuss it explicitly because it is the simplest GVBS
model with more than one ground state.

\noindent
{\bf Example 1. The Majumdar-Ghosh model \cite{MG,Maj}} Nothing new is to
be proved about this model here, as it was already completely analyzed in
\cite{AKLT1}. But still it is a good starter because of its particular
simplicity. The Majumdar-Ghosh model is a spin-1/2 chain with a nnn-interaction
given by:
$$
H_{[0,2]}=h_0=P^{(3/2)}_{012}=\tover23(\es_1\cdot\es_2+\es_2\cdot\es_3
+\es_1\cdot\es_3)+\tover12\idty
$$
where $P^{(3/2)}_{012}$ is the orthogonal projection onto the subspace
with total spin equal to 3/2. At $T=0$ this model breaks the translation
invariance
of the chain and has two pure infinite-volume ground states which are fully
dimerized: $\om_1$ is a product of singlet states on nn pairs of the form
$\{2i,2i+1\}$ and $\om_2$ is obtained from $\om_1$ by translation over one
lattice spacing.  It is quite obvious that $\om_1$ and $\om_2$ are GVBS
states. For a GVBS description of the unique translation invariant ground
state $\tover12(\om_1+\om_2)$ see \cite{FNW1}, p472, Example 6. In order
to check the conditions of Theorem 1.2 it is convenient to consider the model
on a regrouped chain where the new sites are now formed by nn pairs of sites
of the original chain. For concreteness put
$\tilde\A_i=\A_{2i}\otimes\A_{2i+1}$
and we use $\tilde{\phantom{a}}$ to indicate any object related to the
regrouped chain.
Any interval of the regrouped chain corresponds to an interval of even length
of the original chain, where it is easy to see that:
$$
\G^{(1)}_{[0, 2N+1]}=\Cx(\bigotimes_{i=0}^N \phi_{2i,2i+1})
$$
where the superscript $\phantom{a}^{(1)}$ refers to $\om_1$ and
$\phi_{2i,2i+1}$ is
the single state on the nn pair $\{2i,2i+1\}$. For $\om_2$ we have
$$
\G^{(2)}_{[0, 2N+1]}=\{\alpha\otimes\bigotimes_{i=0}^{N-1} \phi_{2i+1,2i+2}
\otimes\beta\mid \alpha,\beta\in\Cx^2\}
$$
So, the local support spaces $\tilde\G_{[M,N]}=\G^{(1)}_{[2M,2N+1]}
\vee\G^{(2)}_{[2M,2N+1]}$ , $N>M$, are 5-dimensional and it was shown
explicitly in \cite{AKLT1}, that
$$
\ker H_{[2M,2N+1]}=\tilde\G_{[M,N]}
$$
Therefore all conditions of Theorem 1.2 are satisfied and the Majumdar-Ghosh
model has a non-vanishing spectral gap in each of its ground states.

\noindent
{\bf Example 1bis: Generalized Majumdar-Ghosh models}
The Majumdar-Ghosh model can be generalized to spins of arbitrary magnitude.
The infinite volume ground states are then the two fully dimerized
states of the spin-$S$ chain, which are just products of the singlet state of
a nearest neighbour pair of spin $S$'s. The Hamiltonian of a GVBS model
with these ground states is clearly not unique. It is also not directly obvious
that it will just contain nearest neighbour and next-nearest neighbour
interactions and nothing else, but the computations of Long and Siak
\cite{LonS2}
show that interactions with further neighbours are, in fact, not needed.
In terms of the total-spin projections $P^{(J)}_{i,j}$ of a pair of spins
of magnitude $S$ at sites $i$ and $j$, the Hamiltonian studied in \cite{LonS2}
reads:
$$
H=-\sum_{i}\{P^{(0)}_{i,i+1} + P^{(0)}_{i+1,i+2}+{1\over 2S+1}\sum_{J=0}^{2S}
(1-(-1)^{2S-J})P^{(J)}_{i,i+2}\}
$$
By adding the constant $(4S+3)/(2S+1)$ to the interaction term,
its lowest eigenvalue can be made to vanish. Any other non-negative
next-nearest-neighbour interaction with the same kernel defines
a GVBS model with exactly the same ground states.
The general results obtained in the present paper also apply to all these
models and demonstrate the existence of a spectral gap above the ground state.
The value of the spectral gap will, of course, depend on the specifics of
the interaction chosen.

\noindent
{\bf Example 2. A ``critical'' VBS-model.}  Our second example is a
one-parameter family of Hamiltonians with nn interaction for a spin-1
chain showing spontaneous $\Ir_2-$symmetry breaking. Let $\lambda\in
[0,1]$ and define
$$
h_i^{(\lambda)}=(\idty-(S^z_i S^z_{i+1})^2)
+\lambda\left(\idty+\es_i\cdot\es_{i+1}-(S^z_i)^2-(S^z_{i+1})^2
+\{\es_i\cdot\es_{i+1},S^z_i S^z_{i+1}\}\right)
$$
where $\{X,Y\}\equiv XY+YX$. In order to study the ground states of
$H^{(\lambda)}_{[M,N]}=\sum_{i=M}^{N-1}h_i^{(\lambda)}$ we first have a
look at the diagonalization of $h^{(\lambda)}$ acting on $\Cx^3\otimes\Cx^3$.
As an orthonormal basis for $\Cx^3$ we take the eigenvectors of $S^z$:
$\ket 1 ,\ket 0 ,\ket -1 $. As a basis for $\Cx^3\otimes\Cx^3$ it is
convenient to define
$$\eqalign{
\xi_1&=\ket 11
,\quad\xi_2=\ket 10 +\ket 01
,\quad\xi_3=\ket 10 -\ket 01   \cr
\xi_4&=\ket 1-1 +\ket -11
,\quad\xi_5=\ket 00
,\quad\xi_6=\ket 1-1 -\ket -11 \cr
\xi_7&=\ket 0-1 -\ket -10
,\quad\xi_8=\ket -10 +\ket 0-1
,\quad\xi_9=\ket -1-1          \cr
}$$
It turns out that the $\xi_j$ are eigenvectors of $h^{(\lambda)}$:
$$
\eqalign{
&h^{(\lambda)}\xi_4 =0 ,\quad       h^{(\lambda)}\xi_6=0\cr
&h^{(\lambda)}\xi_3 = (1-\lambda)\xi_3  ,\quad h^{(\lambda)}\xi_7 =
(1-\lambda)\xi_7\cr
&h^{(\lambda)}\xi_2 = (1+\lambda)\xi_2  ,\quad h^{(\lambda)}\xi_5 = (1+\lambda)
\xi_5,\quad  h^{(\lambda)}\xi_8 = (1+\lambda)\xi_8\cr
&h^{(\lambda)}\xi_1 = 2\lambda\xi_1  ,\quad h^{(\lambda)}\xi_9 =
2\lambda\xi_9\cr
}$$
So, $\spec h^{(\lambda)}=\{0(2), 1-\lambda (2), 1+\lambda (3), 2\lambda(2)\}$
where the numbers between parenthesis denote the degeneracies,
and $h^{(\lambda)}\geq 0$ for all $\lambda\in [0,1]$. But $\ker h^{(\lambda)}$
depends upon whether $\lambda=1$ or $\lambda \in (0,1)$ or $\lambda=0$.
It is now straightforward to determine $\ker H^{(\lambda)}_{[M,N]}$ for
all integers $M,N$, $N>M$. Define vectors $\Omega_1,\Omega_2,\Xi_1,\Xi_2
\in (\Cx^3)^{\otimes N}$ by:
$$
\eqalign{
\Omega_1&= \mid1,-1,\ldots,(-1)^{N+1}\rangle\, ,\quad
\Omega_2=F^{\otimes N}\Omega_1\cr
\Xi_1&= \mid 0,-1,1,-1,\ldots,(-1)^{N+1}\rangle
      -\mid -1,0,1,-1,\ldots,(-1)^{N+1}\rangle\cr
     &\quad +\mid -1,1,0,-1,\ldots,(-1)^{N+1}\rangle
     -\cdots (-1)^{N+1}\mid -1,1,\ldots,(-1)^{N+1},0\rangle\cr
\Xi_2&= F^{\otimes N}\Xi_1\cr
}$$
where $F$ is the spin flip defined by $F\ket 1 =\ket -1  ,F\ket 0 =\ket 0 ,
F\ket -1 = \ket 1 $. For $\lambda\in(0,1)$ $\ker H^{(\lambda)}_{[1,N]}$
is 2-dimensional and spanned by $\Omega_1$ and $\Omega_2$. For $\lambda=1$
the kernel is 4-dimensional and spanned by $\Omega_1,\Omega_2,\Xi_1$ and
$\Xi_2$. For $\lambda=0$ the kernel is $2^N$-dimensional and spanned
by all the Ising configurations (\ie all configurations not containing any
zeros). It is of course obvious that the
spectrum of the model with $\lambda=0$ is entirely discrete and that there
is a gap of magnitude 1. But there is  really nothing interesting in this
model and we will not discuss it any further.

The infinite-volume ground states for any $\lambda\in (0,1]$ are just
the ground states of the Ising antiferromagnet, obtained by extending
$\Omega_1$
and $\Omega_2$ to the infinite chain. Call the respective thermodynamic limits
$\omega_1$ and $\omega_2$. The thermodynamic limits of $\Xi_1$ and $\Xi_2$
are convex combinations of $\omega_1$ and $\omega_2$, depending on how
exactly the finite interval is tending to $\Ir$. So, we repeat,
$\om_1$ and $\om_2$ (and the convex combinations of them) are
the only infinite volume ground states for the model at $\lambda=1$. There
is a difference however between the cases $\lambda<1$ and $\lambda=1$:
in the former case  one has have the property $\ker H^{(\lambda)}_{[M,N]}
=\G_{[M,N]}$, where the $\G_{[M,N]}$ are the local support spaces for the
state $\tover12 (\omega_1+\omega_2)$ (spanned by the two antiferromagnetic
Ising
configurations), whereas in the latter case one does not, because $Xi_1$
and $\Xi_2$ are two additional finite volume ground states.
So, if $\lambda <1$, Theorem 1.2 applies and there is a spectral gap
above the ground state. We will now show that in the case $\lambda=1$
there is no gap and we will see that the low-lying excitations are closely
related to the additional finite-volume ground states.

For $k\in\Rl$ and $N\geq 1$ define the operators
$$
X^\alpha_N(k)=\sum_{-N\leq x<y\leq N} e^{ik(x-y)}S^\alpha_x
e^{\{i\pi\sum_{x<z<y}S^\alpha_z\}}S^\alpha_y
$$
where $\alpha\in\{x,y,z\}$ labels the three spin-1 operators. Note the
similarity
with the string order parameter employed by De Nijs en Rommelse in
\cite{DNR} and Kennedy and Tasaki in \cite{KT1}. Denote by
$\Omega_i^M$, $i=1,2$, the extensions of the vectors $\Omega_i$ to the
interval $[-M,M]$. It is then easy to see that for $M\geq N$ and $\alpha=x,y$
$$
\langle\Omega^M_i\mid X^\alpha_N(k)\Omega^M_j\rangle=0
$$
A straightforward computations yields
$$
{\langle\Omega^M_i\mid X^\alpha_N(k)^* H^{(1)}_{[-M,M]} X^\alpha_N(k)
\Omega^M_i\rangle\over
\langle\Omega^M_i\mid X^\alpha_N(k)^*  X^\alpha_N(k)\Omega^M_i\rangle}
=\left({N-\tover13\over N-\tover12}\right){3\over 4N}
+\left({2N-1\over 2N+1}\right)(1-\cos k)
$$
Hence the gap of $H^{(1)}_{[-M,M]}$ is $\Order({1\over M})$ and the infinite
volume model has no gap.

Another gapless GVBS-model is the following.
Let $\lambda\in [0,1]$ and define
$$
h_i^{(\lambda)}=(S^z_i)^2 + (S^z_{i+1})^2 +\lambda\left\{
-\idty-(S^z_i S^z_{i+1})^2
+\tover12\es_i\cdot\es_{i+1}+\tover12(\es_i\cdot\es_{i+1})^2\right\}
$$
Define vectors $\Omega,\Xi_1,\Xi_0,\Xi_{-1}
\in (\Cx^3)^{\otimes N}$ by:
$$
\eqalign{
\Omega&= \mid 0,0,\ldots,0\rangle\cr
\Xi_1&= \mid 1,0,0,\ldots,0\rangle
     -\mid 0,1,0,\ldots,0\rangle
     +\mid 0,0,1,0,\ldots,0\rangle
     +\cdots (-1)^{N+1}\mid 0,0,\ldots,0,1\rangle\cr
\Xi_0&= \sum_{0<x<y<N} (-1)^{y-x}(S^+_x S^-_y-S^-_x S^+_y)\Omega\cr
\Xi_{-1}&= F^{\otimes N}\Xi_1\cr
}$$
where $F$ is the spin flip defined by $F\ket 1 =\ket -1  ,F\ket 0 =\ket 0 ,
F\ket -1 = \ket 1 $. For $\lambda\in[0,1)$ $\ker H^{(\lambda)}_{[1,N]}=
\Cx\Omega$ and for $\lambda=1$ the kernel becomes 4-dimensional and is
spanned by $\Omega,\Xi_1,\Xi_0,\Xi_{-1}$. The only infinite volume ground
state
for any $\lambda\in[0,1]$ is the limit of the pure state determined by
$\Omega$. For $\lambda<1$ there is a gap by the general theorem. For
$\lambda=1$, there is no gap. It is easy to calculate \eg the energy
of a spin-wave polarized in the x-direction. One finds the simple dispersion
relation
$E_k=1-\cos k$.

\noindent
{\bf Example 3. Models with helical symmetry.}
Next we want to sketch briefly how simple models
exhibiting
helical symmetry breaking can be obtained. In one dimension, non-trivial
helicity is as close as one can hope to get to the chiral symmetry breaking
conjectured to occur in higher dimensions
(see \eg \cite{WWZ} ).

\def\aklt{\hbox{\eightrm AKLT}}

We start from the spin-1 model introduced by Affleck, Kennedy, Lieb and
Tasaki \cite{AKLT2}. Its Hamiltonian is
$$
H^{{\aklt}}=\sum_i\{\tover13 +\tover12
\es_i\cdot\es_{i+1}+\tover16(\es_i\cdot\es_{i+1})^2\}
$$
The unique infinite volume ground state $\om_{{\aklt}}$ of the AKLT-model is
the
GVBS-state determined by
\item{-} the dimension of the auxiliary space is $k=2$ and the isometry $V$
is
given by
$$\eqalign{
V\ket {\tover12} &= \sqrt{\tover23}\ket {1,-\tover12} -\sqrt{\tover12}
\ket {0,\tover12} \cr
V\ket {-\tover12} &=
\sqrt{\tover13}\ket {0,-\tover12} -\sqrt{\tover23}\ket {-1,\tover12}
}$$
\item{-} $\E_A(B)=V^* A\otimes B V$ for all $A\in\M_3,
B\in\M_2$
\item{-} $\om_{{\aklt}}(A_1\otimes\cdots\otimes
A_n)=\tover12\Tr\E_{A_1}\circ\cdots\circ\E_{A_n}
(\idty)$

Our aim is to perturb $\om_{{\aklt}}$ in such a way that part
of the rotation invariance is broken and a state with
non-trivial helicity is obtained. It is then straightforward to construct a
Hamiltonian which has two ground states with opposite helicity. Such a
model can
also be considered as an example of a model where the reflection symmetry
of the
chain is spontaneously broken.

We will denote by $S^x,S^y,S^z$ the usual spin-1 matrices and by
$J^x,J^y,J^z$
the spin-$\tover12$ matrices,
which generate the 3-dimensional irreducible representation $D^{(1)}$ and
the
2-dimensional irreducible representation $D^{(\tover12)}$ of SU(2)
respectively. The isometry $V$ intertwines the representations
$D^{(1)}\otimes
D^{(\tover12)}$ and $D^{(\tover12)}$ and hence
$$
\E_A(D^{(\tover12)}(g) B D^{(\tover12)}(g)^*)=
D^{(\tover12)}(g) \E_{D^{(1)}(g)^* A D^{(1)}(g)}(B) D^{(\tover12)}(g)^*
\deqno(6.1)$$
For $\alpha\in[0,4\pi)$ define:
$$
U(\alpha)=e^{i\alpha J^z}=\pmatrix{e^{i\alpha/2} & 0 \cr
0 & e^{-i\alpha/2}\cr}
$$
and put
$$
\E^{(\alpha)}_A(B)=\E_A(U(\alpha)BU(\alpha)^*)
$$
A family of new GVBS-states is defined by
$$
\om_\alpha(A_1\otimes\cdots\otimes A_n)
=\tover12 \Tr
\E^{(\alpha)}_{A_1}\circ\cdots\circ\E^{(\alpha)}_{A_n}
(\idty)
$$
Using (6.1) it  is straightforward to check the following relation
between $\om_{{\aklt}}$ and $\om_\alpha$:
$$\eqalign{
&\om_\alpha(A_1\otimes\cdots\otimes A_n)
=\cr
&\om_{{\aklt}}(R((n-1)\alpha)^*A_1 R((n-1)\alpha)\otimes R((n-2)\alpha)^*A_1
R((n-2)\alpha)\otimes\cdots\otimes
A_n)
}$$
where $R(\beta)=\exp{i\beta S^z}$ for all $\beta\in \Rl$.
So, $\om_\alpha$ is obtained from $\om_{{\aklt}}$ by a
``twist'' about the $z$-axis over an angle $\alpha$
per lattice spacing.  Obviously $\om_0=\om_{2\pi}=\om_{{\aklt}}$.
(6.1) expresses the rotation invariance of $\om_{{\aklt}}$ and implies that
the state $\om_\alpha$ is still translation invariant.  However the
rotation invariance is reduced to rotations about the $z$-axis only.  The
$\om_\alpha$ are the unique ground states of a family of Hamiltonians
obtained from $H^{{\aklt}}$ by the corresponding ``twist''.  More
interesting is the fact that from Theorem 1.1 it follows that, if
$\alpha\neq -\alpha \bmod 2\pi$, we can also find a finite range
interaction, say $h^{(\alpha)}$, such that the corresponding model has
exactly two ground states: $\om_\alpha$ and $\om_{-\alpha}$.
It is in
principle absolutely straightforward to obtain explicit expressions for the
interactions $h^{(\alpha)}$ by computing the local support spaces of the
states $\om_\alpha$ and by determining a value of $m_0$ for which the
intersection property (2.10) holds.  The actual computation might be
somewhat tedious and not particularly enlightening.
The result is a model of the form
$$
H^{(\alpha)}=\sum_i (\idty-G^{(\alpha)}_{[1+i,m_0+i]})
$$
Theorem 1.2 implies the existence of a spectral gap above the
ground state.

Let us end the discussion of the states $\om_\alpha$ by
computing the various order parameters and correlation functions that are
usually employed to reveal the structure of quantum spin states, in
particular the ones that were investigated in the recent literature on spin-1
chains \cite{DNR,AKLT1,KT1}. We define as usual $J^{\pm}=J^x\pm i
J^y$.
All expectation values in $\om_\alpha$, can be calculated using
\eq(3.5) and Table 1. which fully describes
the operator $\E^{(\alpha)}$.

\item{1)} The magnitization vanishes:
$$
\om_\alpha(S^\gamma_0)=0 \quad \hbox{for } \gamma=x,y,z
$$
\item{2)} The spin-spin correlation functions are periodic but not
necessarily
commensurate with the lattice:
$$\eqalign{
&\om_\alpha(S^z_0 S^\gamma_r)=\om_\alpha(S^\gamma_0 S^z_r)
=\delta_{\gamma,z}\tover43(-\tover13)^r,
\quad r\geq 1\cr
&\om_\alpha(S^x_0 S^x_r)=\om_\alpha(S^y_0S^y_r)
=\tover43(-\tover13)^r\cos r\alpha,
\quad r\geq 1\cr
&\om_\alpha(S^x_0 S^y_r)=-\om_\alpha(S^y_0 S^x_r)
=(-\tover13)^r\sin r\alpha,
\quad r\geq 1
}$$
\item{3)} The den Nijs-Rommelse string order parameter
according to its original definition in \cite{DNR} is given by:
$$
O^\gamma=\lim_{r\to\infty}\om_\alpha(S^\gamma_0
\prod_{x=1}^{r-1}e^{i\pi S^\gamma_x} S^\gamma_r)
$$
and we have
$$\eqalign{
O^z=O^x=O^y=\tover19     &\quad\hbox{ if }\quad \alpha=
k\pi, k\in\Ir\cr
O^z=\tover19, O^x=O^y=0 &\quad\hbox{ else }
}$$
\item{4)} The helicity of the state can be measured by
the correlation function $\chi^z(r)=S^x_0 S^y_r-S^y_0 S^x_r$. The states
$\om_\alpha$ have short range helicity but no long range helical order.
{}From 2)
it follows that
$\chi(r)=-\tover49 \sin r\alpha\neq 0$ if $\alpha\neq k\pi$, but
$\lim_{r\to\infty}\chi(r)=0$.

\smallskip
{
\newdimen\rowheight
\newdimen\columnwidth

\columnwidth=2cm
\rowheight=1cm

\medskip
\centerline{\hbox 
{
\baselineskip=0pt
\everycr={\noalign{\vrule}}
\valign{
\hbox{\vbox to \rowheight{\hrule width \columnwidth\vfil
\hbox to \columnwidth{\hskip 3pt # \hfil}\vskip 5pt
\hrule height 2pt width \columnwidth}}
&\hbox{\vbox to \rowheight{\vfil
\hbox to \columnwidth{\hskip 3pt # \hfil}\vskip 5pt
\hrule width \columnwidth}}
&\hbox{\vbox to \rowheight{\vfil
\hbox to \columnwidth{\hskip 3pt # \hfil}\vskip 5pt
\hrule width \columnwidth}}
&\hbox{\vbox to \rowheight{\vfil
\hbox to \columnwidth{\hskip 3pt # \hfil}\vskip 5pt
\hrule width \columnwidth}}
&\hbox{\vbox to \rowheight{\vfil
\hbox to \columnwidth{\hskip 3pt # \hfil}\vskip 5pt
\hrule width \columnwidth}}
\cr
$\E^{(\alpha)}_A(B)$   &$A=\idty$      &$A=S^z$
&$A=S^+$        &$A=S^-$    \crcr
\noalign{\vrule width 2pt}
$B=\idty$       &$\idty$      &$\tover43 J^z$
&$\tover43 J^+$   &$\tover43 J^-$
\cr
$B=J^z$          &$-\tover13 J^z$ &$-\tover13\idty$
&$0$                                         &$0$
\cr
$B=J^+$          &$-\tover13 e^{i\alpha} J^+$ &$0$
&$-\tover23 e^{i\alpha}\idty$  &$0$
\cr
$B=J^-$          &$-\tover13 e^{-i\alpha}J^-$ &$0$
&$0$           &$-\tover23 e^{-i\alpha}\idty$
\cr
}\hfil}}
\smallskip
\centerline{Table 1. Values taken by the bilinear operator
$\E^{(\alpha)}_A(B)$ on the basis of spin matrices.}}
\smallskip

\noindent
{\bf Example 4.  A model with charge conjugation symmetry breaking.}
Finally we consider a model with spontaneous breaking of the \cc\ symmetry
that was first presented by Affleck, Arovas, Marston and Rabson in
\cite{AAMR}.
We will call it the AAMR-model.  As the construction of the model and the
analysis of its properties is based on the structure of the \irrep s of
SU(4), we have to recall some of the basic facts about these first.  For
more information see \eg \cite{Che}.

The \irrep s of SU(4) are labeled by the Young tableaux with three rows,
including the empty tableau (or alternatively the Young tableau consisting
of a single column of four boxes) which stands for the trivial \rep\ or, in
physical terms, the singlet.  The number of boxes in each row are denoted
by integers $\nu_1,\nu_2,\nu_3$, satisfying $\nu_1\geq\nu_2\geq\nu_3\geq
0$, and $[\nu_1,\nu_2,\nu_3]$ is an alternative way to denote a particular
irreducible representation.
Using Robinson's formula \cite{Che} one easily obtains the following
expression for the dimension of an \irrep :
$$
\dim [\nu_1,\nu_2,\nu_3]=
\tover1{12} (\nu_1+3)(\nu_2+2)(\nu_3+1)(\nu_1-\nu_2+1)
(\nu_1-\nu_3+2)(\nu_2-\nu_3+1)
$$
\eg in what follows we will use
$$
\dim \Yngb 1 1 =6,\quad \dim\Yngb 2 2 =20
$$
The decomposition of a tensor product of two \irrep s into a direct sum of
\irrep s is given by the usual rule for multiplying Young tableaux with
four rows and using the equivalence $[\nu_1,\nu_2,\nu_3,\nu_4]\equiv
[\nu_1-\nu_4,\nu_2-\nu_4,\nu_3-\nu_4]$.  We will \eg need:
$$
\Yngb 1 1 \otimes \Yngb1 1
\cong \Yngd 1 1 1 1 \oplus \Yngc 2 1 1 \oplus \Yngb 2 2
\deqno(6.2)$$
SU(4) is a 15-dimensional Lie group, but it is convenient to represent its
Lie algebra as the traceless subalgebra of the Lie algebra of U(4), \ie we
consider generators $S^\alpha_\beta$, $1\leq \alpha,\beta\leq 4$ satisfying
$$
[S^\alpha_\beta,S^\mu_\nu]=\delta^\mu_\beta S^\alpha_\nu
-\delta^\alpha_\nu
S^\mu_\beta
$$
with the constraint that $\Tr S=\sum_\alpha S^\alpha_\alpha =0$ and the
$S^\alpha_\beta$ are chosen such that $(S^\alpha_\beta)^*=S^\beta_\alpha$.
{}From any representation of this Lie algebra, say generated by
$S^\alpha_\beta$, we can obtain another one generated by $S'^\alpha_\beta$
by putting $S'^\alpha_\beta=-S^\beta_\alpha$.  This is the conjugate \rep\
for which we will systematically use primed quantities.  Of course there is
a corresponding conjugation operation for the \irrep s of SU(4) and hence
for the Young tableaux.  It is described by
$[\nu_1,\nu_2,\nu_3]'=[\nu_1,\nu_1-\nu_3,\nu_1-\nu_2]$, \eg :
$$
\Yngb 2 1 '=\Yngc 2 2 1
$$
In general mutually conjugate \rep s are not equivalent
(see \eg\ the above example), but some are:
$$
\Yngb 1 1 '\cong \Yngb 1 1 ,\quad
\Yngc 2 1 1 '\cong \Yngc 2 1 1 ,\quad
\Yngb 2 2 '\cong \Yngb 2 2
$$
and of course
$$
\Yngd 1 1 1 1 ' = \Yngd 1 1 1 1
$$
A self-conjugate \rep\ and its conjugate are isomorphic, but not identical
(except of course for the case of the singlet \rep ) and hence there is a
non-trivial unitary $C$ implementing this isomorphism which is called the
charge conjugation operator.  $C$ is a spontaneously broken symmetry in the
AAMR-model, which we will introduce now.

The one-site Hilbert space of the AAMR-model is $\Cx^6$ on which one lets
SU(4) act by its 6-dimensional \irrep\ $[1,1,0]$.  So, for each pair of
sites the \irrep s that appear are given by the decomposition (6.2).  Let
$P^{(\yngb 2 2 )}$ denote the orthogonal projection onto the subspace of
$\Cx^6\otimes\Cx^6$ supporting the \irrep\ $[2,2,0]$.  The Hamiltonian of
the AAMR-model is then:
$$
H_{[M,N]}=\sum_{i=M}^{N-1} P^{(\yngb 2 2 )}
\deqno(6.3)$$
As all representations in the decomposition (6.2) are self-conjugate and
distinct, it is obvious that the projection operators onto their supports
commute with $C$ and hence the Hamiltonian (6.3) is \cc\ symmetric.  We now
construct two distinct pure ground states for the model, which are both
SU(4)-invariant and related to each other by \cc .  This implies that the
model exhibits spontaneous breaking of the \cc\ symmetry.  These two ground
states are given in \cite{AAMR} in a convenient representation using
fermion operators in four flavors.  As our main purpose here is to see how
the general results of this paper apply to this model, we prefer to give a
more compact definition of these states as GVBS-states.

Consider two isometries $V$ and $V':\Cx^4\to\Cx^6\otimes\Cx^4$ satisfying
the intertwining relations
$$\Yngb 1 1 \otimes\Ynga 1 V = V\Yngc 1 1 1 ,\quad\quad
\Yngb 1 1 \otimes\Yngc 1 1 1 V' = V'\Ynga 1
$$
The decomposition
$$
\Yngb 1 1 \otimes\Ynga 1 \cong
\Yngc 1 1 1 \oplus \Yngb 2 1
$$
and the conjugate of this relation, imply that these isometries exist and
are unique up to a phase.  For all $A\in\M_6$ we then define the
transformations $\E_A$ and $\E'_A$ of $\M_4$ by
$$\eqalign{
\E_A(B)&=V^* B V\cr
\E'_A(B)&=V'^* B V'
}$$
Because in this model not only the \cc\ symmetry but also translation
invariance is spontaneously broken, it is convenient to consider a
regrouped chain, where the new sites consist of pairs of nearest neighbour
sites of the original chain.  Quantities referring to the regrouped chain
will be denoted by $\tilde{\phantom{a}}$.  Obviously any state $\om$ of the
the regrouped chain is a state of the original chain and vice versa.  Two
GVBS-states $\om$ and $\om'$ are defined by:
$$\eqalign{
\om(A_1\otimes\cdots\otimes A_{2n})=
\tover14\Tr\E_{A_1}\circ\E'_{A_2}\circ\cdots
\circ\E_{A_{2n-1}}\circ\E'_{A_{2n}}
(\idty)\cr
\om('A_1\otimes\cdots\otimes A_{2n})=
\tover14\Tr\E'_{A_1}\circ\E_{A_2}\circ\cdots
\circ\E'_{A_{2n-1}}\circ\E'_{A_{2n}}(\idty)
}$$
It is quite obvious that the trace is invariant under $\E_\idty$ and
$\E'_\idty$, and so $\om\circ\tau=\om'$, where $\tau$ is the translation
over one lattice spacing.  It is also evident that $\om$ and $\om'$ are
related to one another by \cc , as $V$ and $V'$ are.  The fact that \cc\
symmetry is broken can also be expressed by a non-vanishing order parameter
(see \cite{AAMR}).  That $\om$ and $\om'$ are ground states of the model
follows from the transformation properties of $\om\rstr\A_{[1,2]}$ and
$\om'\rstr\A_{[1,2]}$.  One readily sees that the support of the density
matrices describing the restriction of the state to a pair of nearest
neighbour points transform as
$$
\Ynga 1 '\otimes \Ynga 1 \quad\hbox{and }\quad \Yngc 1 1 1 '
\otimes \Yngc 1 1 1
$$
which decompose as
$$
\Ynga 1 '\otimes \Ynga 1 \cong \Yngc 1 1 1 '
\otimes \Yngc 1 1 1 \cong
\Yngc 1 1 1 \otimes \Ynga 1 \cong \Yngd 1 1 1 1 \otimes
\Yngc 2 1 1
$$
and so the supports do not contain $[2,2,0]$ and hence $\om(H)=\om'(H)=0$
and $\om$ and $\om'$ are ground states.

We now would like to apply Theorem 1.2 to get the existence of a spectral gap
in the AAMR-model. The condition one has to check is the following:
we have to verify that for some large enough interval $[1,\el]$,
all zero energy vectors of $H_{[1,\el]}$ are in the supports of $\om$
and $\om'$. This is claimed in \cite{AAMR} but we do not have
a complete argument for this property. It would nice to have
effective techniques to check this kind of properties for this and
more general models.

\let\REF\doref
\Acknow
The author would like to thank Ian Affleck, Michael Aizenman, Joe Conlon,
Mark Fannes,
Elliott Lieb, Fabio Martinelli, Hal Tasaki, Reinhard Werner, and
Horng-Tzer Yau for interesting and useful
discussions. \Thm/gapestimate/ is based on a discussion with
Horng-Tzer Yau. This work has benefited from a travel grant of the
Nationaal Fonds voor Wetenschappelijk Onderzoek, Belgium, and
is financially supported in part by NSF Grants \# PHY-8912067
and \# PHY90-19433 A03.

\parskip=0pt\vsize=22.7truecm\voffset=-.5truecm
\refskip=7pt

\REF Aff1 Aff1 \Jref
I. Affleck
"Large-n limit of SU(n) quantum ``spin'' chains"
Phys. Rev. Lett. @54(1985) 966--969

\REF Aff3 Aff3 \Jref
I. Affleck
"Exact results on the dimerization transition
in SU(n) antiferromagnetic chains"
J. Phys.:Condens. Matter @2(1990)405-415

\REF KT2 KT2 \Jref
T.Kennedy and H. Tasaki
"Hidden $Z_2\times Z_2$ symmetry breaking in Haldane gap
antiferromagnets"
Phys. Rev. @B45(1992)304-307

\REF Dag Dag \Jref
E. Dagotto
"The t-J and frustrated Heisenberg model: a status report
on numerical studies"
Int. J. Mod. Phys. @B5(1991)907-935

\REF WWZ WWZ \Jref
X.G. Wen, F. Wilczeck, A. Zee
"Chiral spin states and superconductivity"
Phys. Rev. @B39(1989)11413-11423

\REF LonS2 LonS2 \Jref
M.W. Long, S. Siak
"An exact solution to a spin-1 chain model"
J. Phys.: Condens. Matter @5(1993)5811-5828

\REF AAMR AAMR \Jref
I. Affleck, D.P. Arovas, J.B. Marston, D.A. Rabson
"SU(2n) Quantum Antiferromagnets with Exact C-Breaking
Ground States"
Nucl. Phys. @B366(1991)467-506

\REF AffH AffH  \Jref
I. Affleck, F.D.M. Haldane
"Critical theory of quantum spin chains"
Phys. Rev. @B36(1987)5291-5300

\REF Aff2 Aff2 \Jref
I. Affleck
"Quantum spin chains and the Haldane gap"
J. Phys.:Condens. Matter @1(1989)3047-3072

\REF SA SA \Jref
E.S. S\o rensen, I. Affleck
"Large-Scale Numerical Evidence for Bose Condensation
in the $S=1$ Antiferromagnetic Chain in a Strong Field"
Phys. Rev. Lett. @71(1993)1633-1636

\REF AKLT2 AKLT2  \Jref
I. Affleck, E.H. Lieb, T. Kennedy, H. Tasaki
"Rigorous results on valence-bond ground states in antiferromagnets"
Phys. Rev. Lett. @59(1987) 799--802

\REF AKLT1 AKLT1 \Jref
I. Affleck, T. Kennedy, E.H. Lieb, H. Tasaki
"Valence bond ground states in isotropic quantum
antiferromagnets"
Commun. Math. Phys. @115(1988) 477--528

\REF FNW3 FNW3 \Jref
M. Fannes, B. Nachtergaele, R.F. Werner
"Valence bond states on quantum spin chains as ground states with
spectral gap"
J. Phys. A: Math. Gen. @24(1991)L185--L190

\REF FNW1 FNW1 \Jref
M. Fannes, B. Nachtergaele, R.F. Werner
"Finitely Correlated States on Quantum Spin Chains"
Commun. Math. Phys. @144(1992)443-490

\REF AAH AAH \Jref
D.P. Arovas, A. Auerbach, F.D.M. Haldane
"Extended Heisenberg models of antiferromagnetism:
analogies to the fractional quantum Hall effect"
Phys. Rev. Lett. @60(1988)531--534

\REF Kna Kna \Jref
S. Knabe
"Energy gaps and elementary excitations for Certain VBS-Quantum
Antiferromagnets"
J. Stat. Phys. @52(1988)627-638

\REF BJ BJ \Jref
R. Botet, R. Julien
"Ground-state properties of a spin-1 antiferromagnetic chain"
Phys. Rev. @B27(1983)613-615

\REF KBJ KBJ \Jref
M. Kolb, R. Botet, R. Julien
"Comparison of ground-state properties for odd half-integer and
integer spin antiferromagnetic Heisenberg chains"
J. Phys. A: Math. Gen. @16(1983)L673-L677

\REF PB PB \Jref
J.B. Parkinson, J.C. Bonner
"Spin chains in a field: Crossover form quantum to classical
behavior"
Phys. Rev. @B32(1985)4703-4724

\REF NB NB \Jref
M.P. Nightingale, H.W. Bl\"ote
"Gap of the linear spin-1 Heisenberg antiferromagnet: a Monte Carlo
calculation"
Phys. Rev. @B33(1986)659-661

\REF Sol Sol \Jref
J. S\'olyom
"Competing bilinear and biquadratic exchange
couplings in spin-1 Heisenberg chains"
Phys. Rev. @B36(1987)8642-8648

\REF CAHS CAHS \Jref
K. Chang, I. Affleck, G.W. Hayden, Z. G. Soos
"A study of the bilinear-biquadratic spin 1 antiferromagnetic chain
using the valence-bond basis"
J.Phys @C1(1989) 153--167

\REF Ken1 Ken1 \Jref
T. Kennedy
"Exact diagonalisations of open spin-1 chains"
J. Phys.: Cond. Matter @2(1990)5737-5745

\REF WH WH \Jref
S.R. White, D.A. Huse
"Numerical Renormalization Group Study of Low-lying
Eigenstates of the Antiferromagnetic S=1 Heisenberg Chain"
Phys. Rev. @B48(1993)3844-3852

\REF Whi1 Whi1 \Jref
S.R. White
"Density Matrix Formulation for Quantum Renormalization Groups"
Phys. Rev. Lett. @69(1992)2863-2866

\REF Whi2 Whi2 \Jref
S.R. White
"Density Matrix Formulation for Quantum Renormalization Groups"
Phys. Rev. @B48(1993)10345-10456

\REF Ken2 Ken2 \Gref
T. Kennedy
"Nonpositive matrix elements for Hamiltonians of spin 1 chains"
preprint

\REF CM1 CM1 \Jref
W.J. Caspers, W. Magnus
"Some exact excited states in a linear antiferromagnetic spin
system"
Phys. Lett. @88A(1982) 103-105

\REF SS SS \Jref
B.S. Shastry, B. Sutherland
"Excitation Spectrum of a Dimerized Next-Neighbor Antiferromagnetic
Chain"
Phys. Rev. Lett. @47(1981)964-967

\REF Hol Hol \Jref
R. Holley
"Rapid convergence to equilibrium in one-dimensional stochastic
Ising models"
Ann. Prob. @13(1985)72-89

\REF AH AH \Gref
M. Aizenman and R. Holley
"Rapid convergence to equilibrium of stochastic Ising models
in the Dobrushin-Shlosman regime"
in "Proceedings of the IMA workshop on percolation and infinite
particle systems", Springer Verlag, Berlin 1987, pp 1-11

\REF LY LY \Gref
S.-L. Lu and H.-T. Yau
"Spectral gap and logarithmic Sobolev Inequality for Kawasaki
and Glauber dynamics"
preprint

\REF MO  MO \Gref
F. Martinelli, E. Olivieri
"Finite volume mixing conditions for lattice spin systems
and exponential approach to equilibrium of Glauber dynamics"
I and II, preprints

\REF Hol2 Hol2 \Jref
R. Holley
"Rapid convergence to equilibrium in ferromagnetic stochastic
Ising models"
Resenhas IME-USP @1(1993)131-149

\REF Hal2 Hal2 \Gref
F.D.M. Haldane
"The Hierarchy of Fractional States and Numerical Studies"
pp 303-352 \inPr
R.E. Prange, S.M. Girvin
"The Quantum Hall Effect"
Springer Verlag, New York 1987

\REF Fro Fro \Jref
J. Fr\"ohlich, U.M. Studer
"U(1)$\times$SU(2) gauge invariance of non-relativistic quantum
mechanics, and generalized Hall effects"
Commun. Math. Phys. @148(1992)553-600

\REF FNW2 FNW2 \Jref
M. Fannes, B. Nachtergaele, R.F. Werner
"Exact Ground States of Quantum Spin Chains"
Europhys. Lett. @10(1989)633--637

\REF Wer1 Wer1 \Jref
R.F. Werner
"Remarks on a quantum state extension problem"
Lett. Math. Phys. @19(1990)319-326

\REF Hal Hal \Jref
F.D.M. Haldane
"Continuum dynamics of the 1-D Heisenberg antiferromagnet:
identification with the O(3) nonlinear sigma model"
Phys. Lett. @93A(1983)464--468

\REF AL AL \Jref
I. Affleck, E.H. Lieb
"A proof of part of Haldane's conjecture on quantum spin chains"
Lett. Math. Phys. @12(1986) 57--69

\REF Maj  Maj \Jref
C.K. Majumdar
"Antiferromagnetic model with known ground state"
J. Phys. C: Cond Matt. @3(1970)911--915

\REF MG  MG \Jref
C.K. Majumdar, D.K. Ghosh
"On next nearest-neighbor interaction in linear chain, I and II"
J. Math. Phys. @10(1969) 1388--1398, and 1399--1402

\REF Kle1 Kle1 \Jref
D.J. Klein
"Variational localized-site cluster expansions.
IX. Many-body valence-bond theory"
Phys. Rev. @B19(1979)870-876

\REF vdB vdB \Jref
P.M. van den Broeck
"Exact value of the ground state energy of the linear
antiferromagnetic Heisenberg chain  with nearest and
next-nearest neighbor interactions"
Phys. Lett. @77A(1980)261-262

\REF Kle2 Kle2 \Jref
D.J. Klein
"Exact ground states for a class of antiferromagnetic
Heisenberg models with short range interactions"
J. Phys. A: Math. Gen. @15(1982)661-671

\REF Cas Cas \Jref
W.J. Caspers
"Exact ground states for a class of linear antiferromagnetic
spin systems"
Physica @115A(1982) 275-280

\REF CM2 CM2 \Jref
W.J. Caspers, W. Magnus
"Exact ground states for a class of linear quantum spin systems"
Physica @119A(1983) 291-294

\REF FNW5 FNW5 \Jref
M. Fannes, B. Nachtergaele, R.F. Werner
"Entropy Estimates for Finitely Correlated States"
Ann. Inst. H. Poincar\'e @57(1992)259-277

\REF Bos1 Bos1 \Jref
I. Bose
"Exact ground and excited states of an antiferromagnetic
quantum spin model"
J. Phys.: Condens. Matt. @1(1989) 9267-9271

\REF KSZ2 KSZ2 \Jref
A. Kl\"umper, A. Schadschneider, J. Zittartz
"Ground state properties of a generalized VBS-model"
Z. Phys. B-Condensed Matter @87(1992) 281-287

\REF KLT KLT \Jref
T. Kennedy, E.H. Lieb, H. Tasaki
"A two-dimensional isotropic quantum antiferromagnet with unique
disordered ground state"
J. Stat. Phys. @53(1988)383-415

\REF CCK CCK \Jref
J. Chayes, L. Chayes, S. Kivelson
"Valence bond ground states in a frustrated two-dimensional
spin 1/2 Heisenberg antiferromagnet"
Commun.Math.Phys. @123(1989)53-83

\REF KK KK \Jref
A.N. Kirillov, V.E. Korepin
"The resonating valence bond in quasicrystals"
Lenin\-grad Math. J. @1(1990)343-377

\REF LonS1 LonS1 \Jref
M.W. Long, S. Siak
"An exactly soluble two-dimensional quantum mechanical Heisenberg model:
quantum fluctuations versus magnetic order"
J. Phys.: Condens. Matter @2(1990)10321-10341

\REF Bos2 Bos2 \Jref
I. Bose
"Two-dimensional spin models with resonating valence
bond ground states"
J. Phys.: Consdens. Matter @2(1990)5479-5482

\REF Bos3 Bos3 \Jref
I. Bose
"Frustrated spin $\tover12$-model in two dimensions with a known
ground state"
Phys. Rev @B44(1991)443-445

\REF Bos4 Bos4 \Jref
I. Bose
"Antiferromagnetic spin models in two dimensions with known
ground states"
Phys. Rev @B45(1992) 13072-13075

\REF FMH1 FMH1 \Jref
W.-D. Freitag, E. M\"uller-Hartmann
"Complete analysis of two-spin correlations of valence bond
solid chains for all integer spins"
Z. Phys. B Condensed Matter @83(1991)381-390

\REF FMH2 FMH2 \Jref
W.-D. Freitag, E. M\"uller-Hartmann
"Spin correlations of inhomogeneous valence bond
solid chains"
Z. Phys. B Condensed Matter @88(1992)279-282

\REF Acc Acc \Jref
L. Accardi
"Topics in Quantum Probability"
Physics Rep. @77(1981) 169-192

\REF AF AF \Jref
L. Accardi, and A. Frigerio
"Markovian Cocycles"
\hfill\break Proc. R. Ir. Acad. @83A{(2)}(1983)251-263

\REF FNW7 FNW7 \Jref
M. Fannes, B. Nachtergaele, R.F. Werner
"Abundance of Translation Invariant Pure
States on Quantum Spin Chains"
Lett. Math. Phys. @25(1992)249-258

\REF FNW6 FNW6 \Jref
M. Fannes, B. Nachtergaele, R.F. Werner
"Finitely correlated pure states"
J.~Funct. Anal. @120(1994)511-534

\REF MS1  MS1  \Jref
F. Monti, A. S\"ut\H o
"Spin 1/2 Heisenberg model on $\Delta$ trees"
Phys. Lett. @156(1991)197-200

\REF MS2  MS2  \Jref
F. Monti, A. S\"ut\H o
"Heisenberg Antiferromagnet on Triangulated Trees"
Helv. Phys. Acta @65(1992)560-595

\REF GW GW \Gref
C.-T. Gottstein, R.F. Werner
"Ground states of the infinite q-deformed Heisenberg ferromagnet"
preprint

\REF Alb1 Alb1 \Jref
C. Albanese
"Unitary Dressing Transformations and Exponential
Decay Below Threshold for Quantum Spin Systems. Part I-II"
Commun. Math. Phys. @134(1990)1-27

\REF Alb2 Alb2 \Jref
C. Albanese
"Unitary Dressing Transformations and Exponential
Decay Below Threshold for Quantum Spin Systems. Part III-IV"
Commun. Math. Phys. @134(1990)237-272

\REF KT1 KT1 \Jref
T.Kennedy and H. Tasaki
"Hidden Symmetry Breaking and the Haldane Phase in S=1 Quantum
Spin Chains"
Commun. Math. Phys. @147(1992)431-484

\REF Mat Mat \Jref
T. Matsui
"Purification and Uniqueness of Quantum Gibbs States"
Commun. Math. Phys. @162(1994)321-332

\REF FF FF \Gref
R. Fernandez, J. Fr\"ohlich
"TBA"
in preparation

\REF KomT KomT \Jref
T. Koma, H. Tasaki
"Symmetry Breaking and Finite-Size Effects
in Quantum Many-Body Systems"
J. Stat. Phys. @76(1994)745-803

\REF Wer2 Wer2 \Gref
R.F. Werner
"Finitely correlated pure states"
in: M. Fannes, C. Maes, and A. Verbeure (eds),
On three levels; micro-, meso, and macro-approaches in physics,
Plenum, New York 1994

\REF HP HP \Gref
F. Hiai, D. Petz
"Entropy Densities for Algebraic States"
preprint

\REF MP MP \Jref
E. Olivieri, P. Picco
"Cluster Expansion for d-Dimensional
Lattice Systems and Finite-Volume Factorization
Properties"
J. Stat. Phys. @59(1990)221-256

\REF LKZ LKZ \Gref
C. Lange, A. Kl\"umper, J. Zittartz
"Exact ground states for antiferromagnetic spin-one chains with nearest
and next-nearest neighbour interactions"
preprint

\REF DNR DNR \Jref
M. den Nijs, K. Rommelse
"Preroughening transitions in crystal surfaces and valence-bond
phases in quantum spin chains"
Phys. Rev. @B40(1989)4709

\REF Che Che \Bref
J.-Q. Chen
"Group Representation Theory for Physicists"
World Scientific, Singapore, 1989

\vfill\eject
\bye